\documentclass[runningheads]{llncs}
\usepackage[T1]{fontenc}
\usepackage{graphicx}
\usepackage{booktabs}
\usepackage[misc]{ifsym}

\usepackage{mwe}

\usepackage{aux}
\begin{document}

\title{Collaborative Representations for Targeted Causal Inference under Outcome Selection}

\titlerunning{Collaborative Representations for Targeted Causal Inference}

\author{Johan de Aguas\inst{}}

\authorrunning{Johan de Aguas}

\institute{Integreat -- Norwegian Centre for Knowledge-driven Machine Learning, Department of Mathematics, University of Oslo, Norway \email{johanmd@math.uio.no}}

\maketitle              

\begin{abstract}
Estimating causal effects with missing outcomes requires learning outcome, exposure, and selection models. Flexible propensity learners can predict treatment or missingness well while worsening target estimation by emphasizing instruments, weak-overlap regions, or variation unrelated to outcome-regression bias. We propose a collaborative representation-based targeted learning method for recovered average treatment effects under outcome selection. The method learns low-dimensional exposure and selection representations from cross-fitted pseudo-outcomes that encode outcome-regression drift. A finite candidate library alternates targeted outcome updates with representation updates, and inner cross-validation selects representation complexity, collaboration strength, and regularization using a target-aware risk. We give high-level sufficient conditions for collaborative robustness, consistency, and asymptotic linearity after outer validation-fold targeting. Simulations show finite-sample bias reduction, with explicit bias--variance trade-offs across configurations and relative to competing estimators.
\end{abstract}

\section{Introduction}

The problem of recovering and estimating causal effects from observational data in the presence of both confounding and missing data has received sustained attention \cite{Mohan2013,Mohan2014,saadati2019adjustment,Bhattacharya2020,NabiFull,mgraphs,NabiSemi}. For missing outcomes, the challenge is twofold: one must account for the endogenous treatment assignment and, at the same time, address the selective observation of outcomes due to  attrition \cite{Bareinboim12,attrition,biele2019bias,HuberDW,Huber}. Over the last decade, the literature has developed a broad collection of recovery tools for this setting, including sufficient graphical conditions \cite{Mohan2014,Correa_Tian_Bareinboim_2018,saadati2019adjustment}, general identification algorithms \cite{Bhattacharya2020}, and search-based procedures \cite{tikka2021causal}. In certain cases, recovered causal effects admit inverse-probability weighted (IPW) representations \cite{HuberDW}; in others, recovery may proceed through generalized adjustment formulas \cite{Correa_Tian_Bareinboim_2018} or through sequential regressions  \cite{ours}. 

To estimate the recovered causal effect, semiparametric estimators are attractive because they permit flexible machine-learning fitting of nuisance functions while retaining asymptotic guarantees and valid inference \cite{MLDR, gruber2009targeted, vanderLaanRubin2006, DML}. These estimators may enjoy double robustness or, in the sequential-regression setting, multiple robustness across distinct patterns of correctly specified nuisance components \cite{multiplyDR,ours, Wei2022,dw, mcisaac2017,AIPW2017}. This robustness, however, is an asymptotic property. In finite samples, the quality of the estimator depends strongly on how the nuisance functions are fitted. In particular, exposure and selection propensity models that are optimized only for predictive accuracy need not optimize the bias--variance trade-off of the target estimand \cite{schnitzer2016variable,nuisanceModels,collaborative2010}. Target-agnostic, prediction-oriented propensity score learners can exacerbate practical positivity issues, yield unstable IPW, and place excessive emphasis on variables that strongly predict treatment or missingness but contribute little to reducing outcome-regression bias, thereby degrading finite-sample performance \cite{instability,seaman2013review,stableIPW}. 


\paragraph{Motivation:}


Two existing lines of work are particularly relevant. The first is \textit{collaborative targeted minimum loss estimation} (CTMLE) \cite{collaborative2010}, which estimates propensity scores in a way that is explicitly tailored to the target parameter, rather than optimized for nuisance prediction alone. By aligning propensity estimation with the residual bias of the outcome regression, CTMLE can improve finite-sample performance while retaining desirable asymptotic properties. In settings where reducing the effective covariate dimension is beneficial, greedy CTMLE implements this idea through stepwise construction of candidate propensity models, adding covariates along a search path and selecting among candidates using a target-aware criterion. However, this procedure can be computationally expensive, and scalable versions often introduce dependence on arbitrary covariate orderings and other search heuristics \cite{scallable}. These limitations are likely to be even more pronounced under attrition, where collaboration must be coordinated across both the treatment-assignment and selection mechanisms.

A second relevant line of work is \textit{Riesz regression} and related methods \cite{chernozhukov2022debiased,chernozhukov2022riesznet}. These approaches bypass explicit propensity-score estimation by directly learning the Riesz representer associated with the target functional, often together with a calibration step \cite{Lars1,Lars3}. This perspective is appealing because it is target-aware from the outset. Yet, its practical success depends on the choice of function class and regularization \cite{chernozhukov2022debiased}. Under weak overlap, the representer-learning problem may also become ill-conditioned, leading to unstable weights and vulnerability to pathological overfitting:  density-ratio representers may reduce the training objective by placing disproportionate mass on vanishing-probability regions (``train-loss hacking'') or by exploiting low-density covariate pockets that are only weakly supported by the data (the ``density chasm'') \cite{chasm,KatoHack,kato2026riesz}.


\paragraph{Contribution:}

We take inspiration from both CTMLE and Riesz regression. Our goal is to replace stepwise collaborative propensity search with a learned low-dimensional representation that remains stable and directly integrated into a plug-in semiparametric estimator for the recovered causal effect. Instead of traversing a sequence of propensity-score models over subsets or orderings of covariates, we parameterize the treatment and selection propensities through low-dimensional learned representations feeding calibrated output heads. Learning low-dimensional representations for causal estimation connects our construction to representation-learning approaches for treatment-effect estimation \cite{shalit2017,johansson2022} and to sufficient dimension reduction for causal modeling \cite{luo2017sdr}; the distinguishing feature here is that the representations are shaped by target-level drift rather than by covariate balance or predictive fit alone.

Collaborative learning is driven by cross-fitted pseudo-outcomes that target the drift components of the sequential outcome regressions. These pseudo-outcomes train representations for the treatment and selection mechanisms along directions most relevant for reducing target-level bias. The method searches over a finite library, alternating targeted outcome-regression updates with representation updates for the propensity components. Model complexity, collaboration weights, regularization, and other hyperparameters are chosen by inner cross-validation using a targeted risk that balances estimated drift and influence-function variance. A final outer cross-validated targeting step yields a substitution estimator, while the representation penalties can be interpreted as Riesz-type losses for nuisance-specific drift functionals.

\section{Preliminaries}

We consider an observational setting with binary point-exposure $A\in\{0,1\}$ (where $A=1$ denotes active treatment), baseline covariates $W$, an additional covariate set $Z$ that may include post-treatment variables, and an outcome $Y$ that is only partially observed. We observe a selection/missingness indicator $R$ and an outcome proxy $Y^\dagger$, where $Y^\dagger=Y$ if $R=1$ and $Y^\dagger=\emptyset$ if $R=0$.

The \textit{average treatment effect} (ATE) is a canonical causal estimand. For a binary point-exposure $A$ and a continuous outcome $Y\in\R$, it is defined as
\begin{equation*}
\psi := \Delta_a  \E\left[Y \mid \doo(A=a)\right],
\end{equation*}
where $\Delta_a$ denotes the contrast operator, $\Delta_a f(a)=f(1)-f(0)$, and $\doo(\cdot)$ is Pearl's intervention operator from the Structural Causal Model framework \cite{PearlCausality}.

If $(W,Z)$ is an $s$-admissible pair relative to the causal $m$-graph of the system, including the missingness mechanism, then the ATE is recoverable from the incomplete-data observational law \cite{ours}. The recovered estimand is then
\begin{align*}
\Psi[P] &:= \E_W\Delta_a Q_2(W,a),\\
Q_2(W,A) &:= \E_{Z\mid W,A}Q_1(W,A,Z),\\
Q_1(W,A,Z) &:= \E\left[Y \mid W,A,Z,R=1\right].
\end{align*}

This representation decomposes the target functional into two nested outcome regressions. The regression $Q_1$ is the \textit{mean-imputation outcome model}, which reconstructs the conditional mean outcome among units with observed outcomes. The regression $Q_2$ is the \textit{meta-regression}, which transports this information back to the treatment level conditional on $W$ \cite{ours}. Given fitted nuisance regressions $\widehat Q=(\widehat Q_1,\widehat Q_2)$, a plug-in estimator of $\psi$ is
$$
\widehat{\psi}
=
\frac{1}{n}\sum_{i=1}^n
\Big\{
\widehat Q_2(W_i,1)-\widehat Q_2(W_i,0)
\Big\},
$$

Under standard regularity and positivity conditions, the recovered ATE functional $\Psi[P]$ is pathwise differentiable in the nonparametric model for the observed-data law $P$. Its \textit{efficient influence function} (EIF) \cite{ours,Huber}, evaluated at $P$ and at an observation $O_i= (W_i,A_i,Z_i,R_i,Y_i^\dagger)\sim P$, can be written as
\begin{align*}
    D(\eta,\psi;O_i)
    &:= D^Y(\eta;O_i)+D^Z(\eta;O_i)+D^W(\eta,\psi;O_i),\\ \nonumber
    D^Y(\eta;O_i)
    &:= \alpha_1(W_i,A_i,Z_i,R_i)
    \left\{Y_i^\dagger-Q_1(W_i,A_i,Z_i)\right\},\\ \nonumber
    D^Z(\eta;O_i)
    &:= \alpha_2(W_i,A_i)
    \left\{Q_1(W_i,A_i,Z_i)-Q_2(W_i,A_i)\right\},\\ \nonumber
    D^W(\eta,\psi;O_i)
    &:= \Delta_a Q_2(W_i,a)-\psi,
\end{align*}
where $\eta=(Q_1,Q_2,\pi_A,\pi_R)$ collects the nuisance functions. The functions $\alpha_1$ and $\alpha_2$ are the corresponding \emph{clever functions/covariates}; equivalently, they act as \emph{Riesz representers} for the linearized contributions of the $Q_1$ and $Q_2$ regressions to the recovered target functional. They are given in terms of the exposure propensity score $\pi_A(W):=\pr(A=1\mid W)$ and the selection propensity score $\pi_R(W,A,Z):=\pr(R=1\mid W,A,Z)$ by
\begin{equation*}
\alpha_2(W_i,A_i)
:=
\frac{A_i-\pi_A(W_i)}
{\pi_A(W_i)[1-\pi_A(W_i)]},
\qquad
\alpha_1(W_i,A_i,Z_i,R_i)
:=
\frac{R_i \alpha_2(W_i,A_i)}
{\pi_R(W_i,A_i,Z_i)}.
\end{equation*}

When the nuisance functions are evaluated at the true observed-data law $P^\circ$, one gets $\E^\circ D(\eta^\circ,\psi^\circ;O)=0$, so the empirical analogue can be used as an estimating equation for $\psi$. This estimating equation is \textit{multiply robust}: it yields consistency when any of the following nuisance components are correctly specified: \textit{(i)} $Q_1$ and $Q_2$, \textit{(ii)} $\pi_A$ and $\pi_R$, or \textit{(iii)} $\pi_A$ and $Q_1$ \cite{ours}. Estimators based on this EIF \cite{ours,Huber} exploit the fact that, in large samples, sufficiently fast nuisance convergence can yield asymptotic linearity. In finite samples, however, performance depends strongly on how the nuisance functions are estimated.


\section{Collaborative robustness}

A central insight of collaborative estimation is that nuisance models should not be optimized solely for predictive accuracy. For the target parameter $\psi$, the key issue is not whether $\pi_A$ or $\pi_R$ are estimated accurately over the entire covariate space, but whether they are estimated accurately in regions where misspecification of $Q_1$ and $Q_2$ would otherwise generate drift in the target estimator \cite{collaborative2010,schnitzer2016variable}.

\begin{theorem}[Collaborative robustness]\label{theo1} Let $\widetilde{\eta}:=(\widetilde{Q}_1,\widetilde{Q}_2,\widetilde{\pi}_A,\widetilde{\pi}_R)$ denote putative nuisance functions, and let $\eta^\circ:=(Q_1^\circ,Q_2^\circ,\pi_A^\circ,\pi_R^\circ)$ and $P^\circ$ denote the true nuisance functions and law. Write the arm-specific treatment drifts $\delta_a$ and the first-stage outcome drift $\varepsilon$
\begin{align*}
    \delta_a(W) &:=\widetilde Q_2(W,a)-Q_2^\circ(W,a),\qquad a\in\{0,1\},\\
    \varepsilon(X) &:=\widetilde Q_1(X)-Q_1^\circ(X),\qquad X=(W,A,Z).
\end{align*}
Assume positivity for the true and the putative propensity scores and bounded second moments. Then the moment equation $\E^\circ D(\widetilde\eta,\psi^\circ;O)=0$ remains a valid estimating equation for $\psi$ if the propensity components are specified collaboratively with the outcome-regression errors in the following sufficient sense:
\begin{align}
\widetilde\pi_A(W)
&=\pr^\circ\!\left(A=1\mid \delta_1(W),\delta_0(W)\right),\label{eqCol1}\\
\widetilde\pi_R(X)
&=\pr^\circ\!\left(R=1\mid W,A,\varepsilon(X)\right).\label{eqCol2}
\end{align}
\end{theorem}

Assumptions and proof are given in the supplementary material.

\Cref{theo1} shows that, even when $\widetilde{Q}_1$ and $\widetilde{Q}_2$ are misspecified, unbiasedness of the EIF estimating equation can still be preserved through collaborative propensity modeling. The propensity components need not recover the full true mechanisms in all directions of the covariate space. Instead, they need to be accurate along the nuisance-error directions that enter the target-level drift. For the exposure component, this requires the two arm-specific drifts $\delta_1$ and $\delta_0$ in general. For the selection component, the selection propensity must align with the first-stage regression drift conditionally on $(W,A)$, because the corresponding EIF term is multiplied by the clever covariate $\alpha_2(W,A)$. 

In the no-attrition case of classical CTMLE \cite{collaborative2010}, the target depends only on the regression $Q_2$, so the exposure-side condition in \eqref{eqCol1} is the relevant one. If the arm-specific drifts $(\delta_1,\delta_0)$ depend only on a proper subset of the covariates in $W$, then fitting the treatment propensity over the full covariate set is unnecessary for bias correction of the target parameter. Searching over reduced or expanded propensity-score models, or penalizing covariates by outcome relevance as in the outcome-adaptive LASSO \cite{shortreed2017oal}, can improve the bias--variance trade-off by avoiding variables that are highly predictive of treatment assignment but largely irrelevant for correcting the target-level drift, and that may otherwise inflate variance through unstable IPW \cite{instability,stableIPW}. These gains are finite-sample rather than purely asymptotic, and they arise precisely because the nuisance-fitting problem is subordinated to the estimation of the target parameter. 

In practice, however, classical CTMLE typically implements this collaboration through stepwise search over candidate propensity-score models. Such search can be computationally expensive, and scalable approximations often reintroduce sensitivity to arbitrary variable orderings and other traversal heuristics \cite{scallable}. Under attrition, the problem is more acute because collaboration must, in principle, be coordinated across two propensity components rather than one. This raises the nontrivial issue of how to specify and justify either two separate search paths or one joint path over candidate variables, and may render the resulting procedure even more sensitive to heuristic design choices. Our proposal addresses this problem by replacing stepwise collaborative search with representation learning guided by cross-fitted \textit{direct noisy proxies} of the drift terms in \eqref{eqCol1} and \eqref{eqCol2}. 


\section{Representation-based CTMLE}\label{sectionfour}

The data consist of samples $O_i=(W_i,A_i,Z_i,R_i,Y_i^\dagger)$, where $Y_i^\dagger$ is only observed when $R_i=1$, and we write $X_i:=(W_i,A_i,Z_i)$. We split the data into a training sample $\mathcal T$ and an evaluation sample $\mathcal E$. In the final implementation, the entire procedure is wrapped in outer $K$-fold cross-fitting, and candidate construction and hyperparameter selection are carried out by inner cross-fitting within $\mathcal T$. To keep notation readable, we suppress inner-fold indices and write all quantities below as if they were computed on a single sample. Thus, whenever pseudo-outcomes or selection losses are evaluated on $\mathcal T$, the corresponding nuisance predictions are understood to be obtained out of fold within $\mathcal T$.

\paragraph{Losses.}
For a given dataset $\mathcal D$ and probabilistic predictors $\pi_A,\pi_R$, define the treatment and attrition cross-entropy losses as
\begin{align*}
\operatorname{CE}_A(\pi_A;\mathcal D)
&=
-\frac{1}{|\mathcal D|}\sum_{i\in\mathcal D}
\Big[
A_i\log \pi_A(W_i)
+
(1-A_i)\log\{1-\pi_A(W_i)\}
\Big],\\
\operatorname{CE}_R(\pi_R;\mathcal D)
&=
-\frac{1}{|\mathcal D|}\sum_{i\in\mathcal D}
\Big[
R_i\log \pi_R(X_i)
+
(1-R_i)\log\{1-\pi_R(X_i)\}
\Big].
\end{align*}
A convenient validation loss for the sequential regressions $Q_1$ and $Q_2$ is
\[
\ell_\omega( Q_1, Q_2;\mathcal D)
=
\frac{1}{|\mathcal D|}\sum_{i\in\mathcal D}
R_i\Big(Y_i^\dagger- Q_1(X_i)\Big)^2
+
\frac{\omega}{|\mathcal D|}\sum_{i\in\mathcal D}
\Big( Q_1(X_i)- Q_2(W_i,A_i)\Big)^2,
\]
where $\omega>0$ is a task weight. The first term assesses the fit of $Q_1$ on observed $Y$, while the second encourages $Q_2$ to track the bridge $Q_1(X)$ through $(W,A)$.

\paragraph{Representations.}
Let $d_A$ and $d_R$ denote the representation dimensions, and define
\[
u_{A,i}=h_{A,\beta_A}(W_i)\in\mathbb R^{d_A},
\qquad
u_{R,i}=h_{R,\beta_R}(X_i)\in\mathbb R^{d_R}.
\]
The treatment propensity is parameterized only through the treatment-drift representation,
\[
\pi_{A,\beta_A,\gamma_A}(W_i)
=
\sigma\{g_{A,\gamma_A}(u_{A,i})\},
\]
whereas the attrition propensity is parameterized with a direct skip for $(W,A)$,
\[
\pi_{R,\beta_R,\gamma_R}(X_i)
=
\sigma\{g_{R,\gamma_R}(W_i,A_i,u_{R,i})\}.
\]
where $\sigma$ is the sigmoid function, $h_{A,\beta_A}$ and $h_{R,\beta_R}$ are deterministic \emph{encoders} into the representation spaces, and $g_{A,\gamma_A}$ and $g_{R,\gamma_R}$ are \emph{output heads} that map these representations to scalars. The encoder $h_{R,\beta_R}$ is asked to preserve the first-stage drift direction $\varepsilon(X)$, while $(W,A)$ are passed directly to the attrition head. In contrast, the treatment condition requires the propensity head to depend on the pair of arm-specific drifts $(\delta_1(W),\delta_0(W))$,  so $h_{A,\beta_A}$ is trained to preserve them.

\paragraph{Pseudo-outcomes.}

The collaborative pseudo-outcomes are constructed so that their conditional means target the oracle drift functions that enter the robustness conditions in \Cref{theo1}. Let $\widetilde\pi_A$, $\widetilde\pi_R$, $\widetilde Q_1$ and $\widetilde Q_2$ denote fixed working nuisance models. For the selection component, define, among units with $R_i=1$,
\[
e_{R,i}
=
\widetilde Q_1(X_i)-Y_i^\dagger .
\]
Since $Q_1^\circ(X)=\E^\circ[Y^\dagger\mid X,R=1]$, this residual satisfies
\[
\E^\circ[e_R\mid X,R=1]
=
\widetilde Q_1(X)-Q_1^\circ(X)
=
\varepsilon(X).
\]
Thus $e_R$ is an observable complete-case noisy proxy for the first-stage drift $\varepsilon(X)$. For the exposure component, first form the selection-corrected pseudo-outcome
\[
\xi_i
=
\widetilde Q_1(X_i)
+
\frac{R_i}{\widetilde\pi_R(X_i)}
\{Y_i^\dagger-\widetilde Q_1(X_i)\}.
\]
We then define the factual-arm residual
\[
e_{A,i}
=
\widetilde Q_2(W_i,A_i)-\xi_i .
\]
When the preceding selection correction is adequate, in particular when
\[
\E^\circ[\xi\mid W,A=a]=Q_2^\circ(W,a),
\]
the conditional mean of this residual satisfies
\[
\E^\circ[e_A\mid W,A=a]
=
\widetilde Q_2(W,a)-Q_2^\circ(W,a)
=
\delta_a(W),
\qquad a\in\{0,1\}.
\]
The exposure representation is therefore trained with two arm-specific auxiliary heads, targeting the two conditional means $\delta_1(W)$ and $\delta_0(W)$. This construction differs from using the EIF residual $Q_1(X)-Q_2(W,A)$, which generally combines first- and second-stage errors. When this correction is adequate, $e_A$ directly targets the arm-specific second-stage drift. The formal verification, including the complete-case nature of the attrition proxy, is given in the supplementary material.


\paragraph{Collaborative and bottleneck penalties.}
Suppressing the candidate index, and treating $e_R$, $e_A$, and the previous-level propensities as fixed, define for any dataset $\mathcal D$ the working treatment weights
\[
\omega_{A,i}^{(1)}
=
\frac{A_i}{\widetilde\pi_A(W_i)},
\qquad
\omega_{A,i}^{(0)}
=
\frac{1-A_i}{1-\widetilde\pi_A(W_i)}.
\]
The collaborative penalties are
\begin{align*}
\mathcal C_R(\beta_R,\kappa_R;\mathcal D)
&=
\frac{1}{|\mathcal D|}\sum_{i\in\mathcal D}
R_i\Big(
 e_{R,i}-f_{R,\kappa_R}\{h_{R,\beta_R}(X_i)\}
\Big)^2,\\
\mathcal C_A(\beta_A,\kappa_A;\mathcal D)
&=
\frac{1}{|\mathcal D|}\sum_{i\in\mathcal D}
\sum_{a\in\{0,1\}}
\omega_{A,i}^{(a)}
\Big(
 e_{A,i}-f_{A,\kappa_A}^{(a)}\{h_{A,\beta_A}(W_i)\}
\Big)^2.
\end{align*}
Here $f_{R,\kappa_R}$ and $f_{A,\kappa_A}=(f_{A,\kappa_A}^{(0)},f_{A,\kappa_A}^{(1)})$ are auxiliary regression heads trained to predict the direct drift proxies from the learned latent codes. The factor $R_i$ restricts the attrition proxy regression to units for which $Y^\dagger$ is observed. The treatment weights use only observations from the corresponding factual arm and, when the working propensity is close to the true propensity, make the arm-specific projection close to marginal in $W$. At the population level, these losses yield the complete-case projection of $\varepsilon(X)$ and weighted projections of $(\delta_1(W),\delta_0(W))$ onto the latent representation classes. These penalties therefore encourage the representations to preserve precisely those directions required by the collaborative robustness conditions in \eqref{eqCol1}--\eqref{eqCol2}. 

We also use surrogate bottleneck penalties
\[
\mathcal B_A(\beta_A;\mathcal D)
=
\frac{1}{|\mathcal D|}\sum_{i\in\mathcal D}\|h_{A,\beta_A}(W_i)\|_2^2,
\qquad
\mathcal B_R(\beta_R;\mathcal D)
=
\frac{1}{|\mathcal D|}\sum_{i\in\mathcal D}\|h_{R,\beta_R}(X_i)\|_2^2.
\]
These penalties act directly on the encoders and yield more compressed and therefore more constrained propensity representations. Because neural encoders and heads are scale-nonidentifiable, the implementation standardizes the latent codes within each training fold and uses ordinary weight decay on the propensity and auxiliary heads; otherwise the encoder penalty could be partially offset by rescaling the downstream heads without changing the fitted scores.

\paragraph{Hyperparameter library and schedule.}
Let $\Theta_n$ denote a finite hyperparameter library. A library element
\[
\theta=(d_A,d_R,\omega,\lambda_A,\lambda_R,\tau_A,\tau_R,\rho)
\in\Theta_n
\]
specifies the representation dimensions, sequential-regression weight, collaboration penalties, bottleneck penalties, and schedule exponent; the maximum budget $B$ is fixed and the selected budget is $b$. For fixed $\theta$ and $b\in\{0,1,\ldots,B\}$, define
\[
\lambda_A^b=\lambda_A\Big(1-\frac{b}{B}\Big)^\rho,
\quad
\quad
\tau_A^b=\tau_A\Big(1-\frac{b}{B}\Big)^\rho,
\]
and equivalently $\lambda_R^b$ and $\tau_R^b$. Hence the terminal candidate for a given $\theta$ has $\lambda_A^{B}=\lambda_R^{B}=\tau_A^{B}=\tau_R^{B}=0$ and reduces to an ordinary non-collaborative cross-entropy propensity fit within the representation class. As $b$ increases, the propensity models become less constrained. The path therefore interpolates between strongly collaborative, strongly bottlenecked candidates and the standard fit.

The final estimator is selected over the joint configuration--budget library
\begin{equation}\label{eq:library}
\mathcal S_n:=\{s=(\theta,b):\theta\in\Theta_n,\ b=1,\ldots,B\},
\qquad N_n:=|\mathcal S_n|,
\end{equation}
with $b=0$ retained only as path initialization.

\paragraph{Targeted selection.}\label{sec:integrated}
The sequential loss $\ell_\omega$ measures regression fit, not directly the remaining target-level drift. Selecting on $\ell_\omega$ alone is therefore only a surrogate: its population minimizer need not minimize the target remainder. We instead use a targeted criterion that estimates the candidate's squared EIF drift plus a second-moment penalty. Let $\mathcal T=\bigcup_{v=1}^{V}\mathcal T_v$ denote the inner folds. For each $s\in\mathcal S_n$, let $\widehat\eta^s_{-v}$ be the candidate built using $\mathcal T\setminus\mathcal T_v$ only, and define the validation estimate
\[
\widehat\psi^s_{-v,\mathcal T_v}
:=
\E_{\mathcal T_v}
\{\widehat Q^{s}_{2,-v}(W,1)-\widehat Q^{s}_{2,-v}(W,0)\}.
\]
The held-out EIF mean and second-moment are
\begin{equation*}
\widehat b_n(s)
=
\frac{1}{V}\sum_{v=1}^{V}
\E_{\mathcal T_v}
D\big(\widehat\eta^{s}_{-v},\widehat\psi^s_{-v,\mathcal T_v};O\big),\quad
\widehat v_n(s)
=
\frac{1}{V}\sum_{v=1}^{V}
\E_{\mathcal T_v}
D\big(\widehat\eta^{s}_{-v},\widehat\psi^s_{-v,\mathcal T_v};O\big)^2.
\end{equation*}
We select using
\begin{equation}\label{eq:targeted-risk}
\mathcal R_n(s)
=
\widehat b_n(s)^2
+
\frac{1}{|\mathcal T|}\widehat v_n(s)
+
\zeta\operatorname{CV}_{\mathrm{in}}\{\ell_\omega(\widehat Q_1^s,\widehat Q_2^s)\},
\end{equation}
where $\zeta\geq0$ is optional. The theoretical default is $\zeta=0$, in which case the criterion targets the target-MSE scale: squared EIF drift plus an influence-function second-moment penalty. A small regression-loss component can improve finite-sample stability, but then the asymptotic argument requires the resulting combined risk to remain locally equivalent to the target remainder.

The selected configuration is
\begin{equation*}
\widehat s=(\widehat\theta,\widehat b)
\in
\arg\min_{s\in\mathcal S_n}
\mathcal R_n(s).
\end{equation*}
This step turns the representation dimension, regularization strengths, and computational budget into components of the estimator rather than externally fixed tuning constants. As in Super Learner and cross-validated TMLE, selection over a finite or slowly growing library can be absorbed into the asymptotic analysis by treating the selected candidate as a data-adaptive element of the candidate library \cite{superlearner2007,cvtmle2011,vdv2006oracle}. A one-standard-error rule may be used as a conservative finite-sample variant, choosing the simpler candidate, smaller $d_A,d_R$, larger regularization, and smaller $b$, among candidates whose validation loss is statistically indistinguishable from the minimum.

\paragraph{Step 0: baseline candidate.}
For each $\theta\in\Theta_n$, begin by fitting initial outcome regressions on $\mathcal T$:
\[
\widehat Q_1^{0}
\approx
\mathbb E[Y\mid X,R=1],
\qquad
\widehat Q_2^{0}
\approx
\mathbb E[\widehat Q_1^{0}(X)\mid W,A].
\]
We then set intercept-only propensity estimators
\[
\widehat\pi_A^{0}(W)
:=
\frac{1}{|\mathcal T|}\sum_{i\in\mathcal T}A_i,
\qquad
\widehat\pi_R^{0}(X)
:=
\frac{1}{|\mathcal T|}\sum_{i\in\mathcal T}R_i.
\]
Using $(\widehat Q_1^{0},\widehat Q_2^{0},\widehat\pi_A^{0},\widehat\pi_R^{0})$ as initial nuisance estimates, we apply the TMLE fluctuation step for the recovered ATE \cite{ours} and obtain
\begin{equation}\label{eq:TMLE1}
(\widehat Q_1^{\star 0},\widehat Q_2^{\star 0})
\leftarrow
\operatorname{TMLE}\Big[
\widehat Q_1^{0},\widehat Q_2^{0},\widehat\pi_A^{0},\widehat\pi_R^{0};\mathcal T
\Big].
\end{equation}
Although the baseline propensity estimators are constant, the associated clever covariates need not vanish, so this fluctuation step may still update the initial regressions. We initialize the budget path with either the targeted tuple $\widehat\eta^{\theta,0}
=
\Big(
\widehat Q_1^{\star 0},
\widehat Q_2^{\star 0},
\widehat\pi_A^{0},
\widehat\pi_R^{0}
\Big)
$ or the untargeted tuple $\widehat\eta^{\theta,0}
=
\Big(
\widehat Q_1^{0},
\widehat Q_2^{0},
\widehat\pi_A^{0},
\widehat\pi_R^{0}
\Big)$. The targeted initialization may stabilize the first update by removing the drift captured by the initial propensity estimates, whereas the untargeted initialization keeps \(b=0\) closer to a plug-in starting point.

\paragraph{Budgeted candidate step $b=1,\dots,B$.}
For each candidate $b$ and fixed hyperparameter value $\theta$, proceed as follows.

\begin{enumerate}
\item Set the current working outcome regressions equal to the previous targeted candidate:
$$
\widehat Q_1^{b}\leftarrow \widehat Q_1^{\star (b-1)},
\qquad
\widehat Q_2^{b}\leftarrow \widehat Q_2^{\star (b-1)}.
$$

\item Construct collaborative pseudo-outcomes on $\mathcal T$. For units with $R_i=1$, define the direct attrition-drift proxy
\[
e_{R,i}^{b}
=
\widehat Q_1^{b}(X_i)-Y_i^\dagger .
\]
Next define the selection-corrected pseudo-outcome
\[
\xi_i^{b}
=
\widehat Q_1^{b}(X_i)
+
\frac{R_i}{\widehat\pi_R^{b-1}(X_i)}
\{Y_i^\dagger-\widehat Q_1^{b}(X_i)\},
\]
where the residual term is set to zero when $R_i=0$. Finally, for all units define the factual-arm second-stage drift proxy
\[
e_{A,i}^{b}
=
\widehat Q_2^{b}(W_i,A_i)-\xi_i^{b}.
\]

\item Fit the candidate attrition and treatment propensity models on $\mathcal T$
\begin{align*}
(\widehat\beta_R^{b},\widehat\gamma_R^{b},\widehat\kappa_R^{b})
&\in
\mathop{\arg \min}\limits_{\beta_R,\gamma_R,\kappa_R}
\Big[
\operatorname{CE}_R(\pi_{R,\beta_R,\gamma_R})
+
\lambda_R^b\,\mathcal C_R(\beta_R,\kappa_R)
+
\tau_R^b\,\mathcal B_R(\beta_R)
\Big],\\
(\widehat\beta_A^{b},\widehat\gamma_A^{b},\widehat\kappa_A^{b})
&\in
\mathop{\arg \min}\limits_{\beta_A,\gamma_A,\kappa_A}
\Big[
\operatorname{CE}_A(\pi_{A,\beta_A,\gamma_A})
+
\lambda_A^b\,\mathcal C_A(\beta_A,\kappa_A)
+
\tau_A^b\,\mathcal B_A(\beta_A)
\Big],
\end{align*}
where $\mathcal C_R$ uses $e_{R,i}^b$, $\mathcal C_A$ uses $e_{A,i}^b$ with the fixed previous-level treatment weights
\[
\omega_{A,i}^{(1),b-1}=
\frac{A_i}{\widehat\pi_A^{b-1}(W_i)},
\qquad
\omega_{A,i}^{(0),b-1}=
\frac{1-A_i}{1-\widehat\pi_A^{b-1}(W_i)}.
\]

\item Form the candidate propensity estimators
\[
\widehat\pi_A^{b}(W_i)
=
\sigma\big(g_{A,\widehat\gamma_A^{b}}\{h_{A,\widehat\beta_A^{b}}(W_i)\}\big),
\qquad 
\widehat\pi_R^{b}(X_i)
=
\sigma\big(g_{R,\widehat\gamma_R^{b}}\{W_i,A_i,h_{R,\widehat\beta_R^{b}}(X_i)\}\big).
\]
In the implementation, we clip them below at the fixed level $\epsilon_{\mathrm{clip}}$ and above at $1-\epsilon_{\mathrm{clip}}$; this clipping level is not a library coordinate.

\item Apply the TMLE targeting step for the recovered ATE, using $\widehat Q_1^{b},\widehat Q_2^{b}$ as initial regressions and $\widehat\pi_A^{b},\widehat\pi_R^{b}$ as candidate propensities:
\begin{equation}\label{eq:TMLE2}
(\widehat Q_1^{\star b},\widehat Q_2^{\star b})
\leftarrow
\operatorname{TMLE}\Big[
\widehat Q_1^{b},\widehat Q_2^{b},\widehat\pi_A^{b},\widehat\pi_R^{b};\mathcal T
\Big].
\end{equation}

\item Store the candidate 
$
\widehat\eta^{\theta,b}
=
\Big(
\widehat Q_1^{\star b},
\widehat Q_2^{\star b},
\widehat\pi_A^{b},
\widehat\pi_R^{b}
\Big)$ 
and continue until $b=B$.
\end{enumerate}

\paragraph{Final estimator.}
The inner cross-validation step selects a joint hyperparameter--budget pair $\widehat s=(\widehat\theta,\widehat b)$. After selection, the candidate construction procedure is rerun on the full outer training sample $\mathcal T$ with the selected pair, yielding training-fold nuisance estimates
$
\overline\eta^{\widehat s}
=
(\overline Q_1,\overline Q_2,\overline\pi_A,\overline\pi_R)
$.
To obtain the fold-specific estimator, we keep the selected representations and propensity scores fixed and apply a final low-dimensional TMLE fluctuation on the held-out evaluation fold $\mathcal E$:
\[
(\widehat Q_1^\sharp,\widehat Q_2^\sharp)
\leftarrow
\operatorname{TMLE}
\Big[
\overline Q_1,
\overline Q_2,
\overline\pi_A,
\overline\pi_R;\mathcal E
\Big].
\]
The fold-specific estimator is then
$$
\widehat{\psi}^\sharp_{\mathcal E}
=
\frac{1}{|\mathcal E|}\sum_{i\in\mathcal E}
\Big
\{
\widehat Q_2^\sharp(W_i,1)-\widehat Q_2^\sharp(W_i,0)
\Big
\}.
$$
Outer cross-fitting averages $\widehat{\psi}^\sharp_{\mathcal E_k}$ across validation folds. This final validation-fold targeting step ensures that the empirical EIF equation is solved on the same observations used to evaluate the fold-specific plug-in estimate. Candidate construction, representation learning, and hyperparameter selection remain separated from $\mathcal E$, while the final targeting step estimates only a low-dimensional fluctuation parameter. If one instead targets only on $\mathcal T$ and evaluates a plug-in estimator on $\mathcal E$, the usual empirical EIF cancellation does not automatically hold on $\mathcal E$, and an additional stability argument is required.

For each fixed $\theta$, the terminal candidate $B$ coincides with the ordinary, non-collaborative propensity fit in that representation class. Including this endpoint allows the procedure to revert adaptively to the standard estimator when collaboration does not improve out-of-sample targeted fit. We emphasize that $b=B$ reduces to the ordinary TMLE of \cite{ours} only when the selected representation class is flexible enough to represent the full propensity mechanisms, for example with sufficiently large $d_A,d_R$ and negligible bottleneck penalty. For moderate $d_A,d_R$, it is a representation-constrained TMLE, which is a deliberate regularization.

\section{Asymptotic validity}

By the standard TMLE fluctuation argument, the targeting steps in \cref{eq:TMLE1,eq:TMLE2} solve the empirical EIF equation on the sample used for targeting, up to numerical optimization error \cite{ours,targetedlearning2011}. In the final estimator, this equation is solved on each outer validation fold after the hyperparameters have been selected on the corresponding training fold:
\[
\E_{\mathcal E}
D(\widehat\eta^{\widehat s,\sharp},\widehat\psi^\sharp_{\mathcal E};O)
=o_{P^\circ}(n^{-1/2}).
\]
This validation-fold targeting is the CV-TMLE analogue of the usual empirical EIF-solving step \cite{cvtmle2011}.

\begin{theorem}[High-level asymptotic validity]\label{thm:main}
Let $\psi^\circ=\Psi[P^\circ]$ be the true recovered ATE. Let $\mathcal S_n$ in \eqref{eq:library} be the joint hyperparameter--budget library, and let $\widehat s\in\arg\min_{s\in\mathcal S_n}\mathcal R_n(s)$ be the inner-cross-validated selector based on \eqref{eq:targeted-risk}. Let $\widehat\eta^{\widehat s,\sharp}$ and $\widehat\psi^\sharp$ denote the outer validation-targeted nuisance tuple and the resulting cross-fitted plug-in estimator. Write $\bar\eta$ for the probability limit of the selected nuisance tuple; under collaborative robustness $\bar\eta$ need not equal $\eta^\circ$. Assume:

\begin{enumerate}
\item[(A1)] \textbf{Sampling, positivity, and boundedness.}
$O_1,\dots,O_n$ are i.i.d.\ from $P^\circ$. There exists $\epsilon_0>0$ such that
\[
\epsilon_0 \le \pi_A^\circ(W) \le 1-\epsilon_0,
\qquad
\epsilon_0 \le \pi_R^\circ(X) \le 1
\]
almost surely. The outcome, nuisance estimators after clipping, and EIFs induced by candidates in $\mathcal S_n$ are uniformly bounded.

\item[(A2)] \textbf{Library-size and oracle-selection control.}
The library is finite for each $n$ and either $N_n$ is fixed or grows slowly enough that the cross-validation oracle inequality applies. A sufficient condition for the empirical-process part of selection is $\log N_n=o(n^{1/2})$, together with the usual boundedness or sub-Gaussian envelope conditions. Equivalently, the selected risk satisfies
\[
\mathcal R_n(\widehat s)-\inf_{s\in\mathcal S_n}\mathcal R_n(s)=o_{P^\circ}(1),
\]
up to numerical optimization error \cite{vdv2006oracle,superlearner2007}.

\item[(A3)] \textbf{Existence of an asymptotically valid oracle configuration.}
There exists a possibly data-dependent sequence $s_n^\dagger\in\mathcal S_n$ such that
\[
\left|\operatorname{Rem}\big(\widehat\eta^{s_n^\dagger,\sharp},\eta^\circ\big)\right|
=
o_{P^\circ}(1),
\qquad
\operatorname{Rem}(\eta,\eta^\circ)
:=
\E_{P^\circ}\!\left[D(\eta,\psi^\circ;O)\right].
\]
This assumption holds when an oracle configuration satisfies one of the multiple-robustness regimes or the collaborative regimes of \Cref{theo1}.

\item[(A4)] \textbf{Selection-risk transfer and stability.}
For any sequence $\widetilde s_n\in\mathcal S_n$, if
\[
\mathcal R_n(\widetilde s_n)-\mathcal R_n(s_n^\dagger)=o_{P^\circ}(1),
\]
then
\[
\left|\operatorname{Rem}\big(\widehat\eta^{\widetilde s_n,\sharp},\eta^\circ\big)\right|
-
\left|\operatorname{Rem}\big(\widehat\eta^{s_n^\dagger,\sharp},\eta^\circ\big)\right|
=
o_{P^\circ}(1).
\]
This condition is plausible when the leading part of $\mathcal R_n$ is the held-out EIF-drift criterion in \eqref{eq:targeted-risk}, because the criterion targets the squared drift. If the optional regression-loss component dominates, or if the procedure is refitted after selection, the condition is a stability assumption transferring the oracle property from the selected library index to the final refitted estimator.

\item[(A5)] \textbf{Cross-fitted empirical-process control.}
With
\[
F(\widehat\eta,\bar\eta,\psi^\circ;O)
=
D(\widehat\eta,\psi^\circ;O)-D(\bar\eta,\psi^\circ;O),
\]
the selected nuisance tuple satisfies, fold by fold,
\[
(\E_{\mathcal E}-\E^\circ)
F(\widehat\eta^{\widehat s,\sharp},\bar\eta,\psi^\circ;O)
=
o_{P^\circ}(n^{-1/2}).
\]
Outer cross-fitting, together with the finite-dimensional validation-fold fluctuation, avoids Donsker restrictions for the initial nuisance learners; the selected-candidate version requires the same control uniformly over the finite library, with the usual $\sqrt{\log N_n}$ max-inequality factor.
\end{enumerate}

Then the following hold.

\begin{enumerate}
\item[(i)] \textbf{Consistency.}
If assumptions \textnormal{(A1)}--\textnormal{(A5)} hold, then
\[
\widehat\psi^\sharp \xrightarrow{P^\circ} \psi^\circ.
\]

\item[(ii)] \textbf{Rate asymptotic linearity.}
If, in addition, the oracle satisfies
\[
\left|\operatorname{Rem}\big(\widehat\eta^{s_n^\dagger,\sharp},\eta^\circ\big)\right|=o_{P^\circ}(n^{-1/2}),
\]
and \textnormal{(A2)}--\textnormal{(A4)} hold at the $o_{P^\circ}(n^{-1/2})$ scale, then
\[
\sqrt n\big(\widehat\psi^\sharp-\psi^\circ\big)
=
\frac{1}{\sqrt n}\sum_{i=1}^n D(\bar\eta,\psi^\circ;O_i)
+
o_{P^\circ}(1)
\rightsquigarrow
N\!\left(0,\Var_{P^\circ}\!\big(D(\bar\eta,\psi^\circ;O)\big)\right).
\]
Thus collaborative robustness can deliver $\sqrt n$-asymptotic normality even when $\bar\eta\neq\eta^\circ$, but the limiting influence function is then generally not the EIF.

\item[(iii)] \textbf{Efficiency.}
If, moreover, all nuisance components are consistent for the truth, $\bar\eta=\eta^\circ$, and the usual second-order product-rate conditions hold, then
\[
\sqrt n\big(\widehat\psi^\sharp-\psi^\circ\big)
\rightsquigarrow
N\!\left(0,\Var_{P^\circ}\!\big(D(\eta^\circ,\psi^\circ;O)\big)\right),
\]
and the estimator attains the semiparametric efficiency bound.
\end{enumerate}
\end{theorem}

Proofs and a remark on the high-level oracle-transfer and stability assumptions are given in the supplementary material. The theorem treats $d_A,d_R$, $b$, $\omega$, $\lambda_A$, $\lambda_R$, $\tau_A$, $\tau_R$, and $\rho$ as selected over a finite or slowly growing library; it does not prove that the neural optimizer consistently discovers an oracle representation.

\subsection{Variance estimation}\label{sec:variance}
For a single outer cross-fitting split, we use the empirical variance of the EIF-form estimating function evaluated at the selected validation-targeted nuisance estimates:
\[
\widehat{\Var}_{\mathrm{IF}}
=
\frac{1}{n}
\widehat{\Var}_n
\!\left[
D(\widehat\eta^{\widehat s,\sharp},\widehat\psi^\sharp;O)
\right].
\]
If the whole procedure is repeated over $M$ independent random split assignments, with estimates $\widehat\psi_m$, EIF-form variance  $\widehat{\Var}_{\mathrm{IF},m}$, and $\bar\psi=M^{-1}\sum_{m=1}^M\widehat\psi_m$, define
\[
V_W=\frac{1}{M}\sum_{m=1}^M \widehat{\Var}_{\mathrm{IF},m},
\qquad
V_B=\frac{1}{M-1}\sum_{m=1}^M(\widehat\psi_m-\bar\psi)^2.
\]
If the reported point estimate is the average $\bar\psi$, the pure Monte Carlo contribution of split randomization is $V_B/M$. A conservative instability-aware interval may instead use $V_W+V_B$, treating split and configuration variability as part of the reported uncertainty. 

\section{Simulation task}\label{sec:simulation}

We evaluate the proposed collaborative representation-based TMLE in a simulation study for the recovered average treatment effect under outcome selection. The design is intentionally challenging. The sequential outcome regressions are deliberately misspecified, which creates a difficult setting for the plug-in estimator. At the same time, the treatment and missingness mechanisms include many strong treatment instruments and missingness instruments. These variables make the propensity scores highly predictable, but much of this predictive signal is not directly relevant for reducing target-parameter bias. As a result, standard one-step and TMLE estimators using aggressively predictive neural propensity scores can suffer from unstable inverse-probability corrections. The simulation therefore probes the central motivation of collaborative learning: propensity score models should be learned with respect to the target parameter, not only with respect to treatment or missingness prediction. Further details on the data-generating process, nuisance learners, collaborative path, and inner targeted selector are given in the supplementary material. Code is available at: \texttt{\url{https://github.com/johandh2o/cr-tmle-outcome-selection}}

The collaborative candidates use two fixed representation/regularization settings  \((d_A,d_R,\omega,\lambda_A,\lambda_R,\tau_A,\tau_R,\rho)\):
\(\theta_1=(3,3,0.6,2.00,5.00,0.015,0.005,1.0)\) and
\(\theta_2=(4,4,0.8,3.00,3.00,0.010,0.010,1.0)\). The library estimator selects between them over budgets \(b\in\{1,\ldots,B\}\), with \(b=0\) used only to initialize the path.

\begin{table}[!htbp]
\centering
\small
\setlength{\tabcolsep}{5pt}
\renewcommand{\arraystretch}{1.08}
\caption{Simulation performance by estimator and sample size. Bias and MSE are rounded to three decimals. Within each sample size, the smallest absolute bias and smallest MSE are shown in bold (average values over 100 Monte Carlo repetitions).}
\label{tab:simulation-results-wide}
\begin{tabular}{lrrrrrr}
\toprule
& \multicolumn{2}{c}{$n=3000$} 
& \multicolumn{2}{c}{$n=5000$} 
& \multicolumn{2}{c}{$n=7000$} \\
\cmidrule(lr){2-3}
\cmidrule(lr){4-5}
\cmidrule(lr){6-7}
Estimator 
& Bias & MSE 
& Bias & MSE 
& Bias & MSE \\
\midrule
CR-TMLE, $\theta_1$ only
& \textbf{-0.477} & 4.787
& \textbf{0.004} & 6.267
& \textbf{0.146} & 3.479 \\

CR-TMLE, $\theta_2$ only
& -0.539 & 5.915
& 0.145 & 4.685
& 0.176 & 2.909 \\

CR-TMLE library, $\theta_1$ and $\theta_2$
& -0.871 & 6.228
& -0.059 & 5.575
& -0.183 & 4.403 \\

One-step, neural propensities
& 1.044 & \textbf{1.230}
& 0.271 & 4.433
& 0.714 & 0.932 \\

TMLE, neural propensities
& -2.055 & 11.492
& -0.824 & 6.665
& -2.134 & 10.819 \\

Plug-in, linear outcome reg.
& 1.531 & 2.428
& 1.520 & 2.371
& 1.521 & 2.364 \\

Ridge Riesz regression
& 1.078 & 1.295
& 0.628 & \textbf{0.575}
& 0.849 & \textbf{0.794} \\
\bottomrule
\end{tabular}
\end{table}

Table~\ref{tab:simulation-results-wide} illustrates distinct failure modes induced by the simulation design. The plug-in estimator has a persistent positive bias, approximately \(1.52\) across all sample sizes, which is consistent with the deliberate misspecification of the sequential outcome regressions. The neural-propensity TMLE benchmark is also unstable in this design. It has negative bias at every sample size, with biases \(-2.055\), \(-0.824\), and \(-2.134\), and it has the largest MSE in all three sample sizes. This behavior suggests that the targeting step can amplify instability when the propensity models devote substantial capacity to treatment instruments and missingness instruments that improve prediction but do not necessarily reduce the drift of the target estimator.

The one-step and ridge Riesz estimators achieve the lowest MSE in these experiments. The one-step estimator has the smallest MSE at \(n=3000\), with MSE \(1.230\), while ridge Riesz has the smallest MSE at \(n=5000\) and \(n=7000\), with MSEs \(0.575\) and \(0.794\), respectively. However, these low-MSE results should be interpreted together with their remaining positive bias. The one-step estimator has biases \(1.044\), \(0.271\), and \(0.714\), while the ridge Riesz estimator has biases \(1.078\), \(0.628\), and \(0.849\) for \(n=3000\), \(5000\), and \(7000\), respectively. Thus, in this simulation, the favorable MSE of these benchmarks appears to be driven partly by lower variability rather than by uniformly accurate bias correction.

The collaborative estimators show a different pattern. The \(\theta_1\)-only CR-TMLE has the smallest absolute bias at all three sample sizes, with biases \(-0.477\), \(0.004\), and \(0.146\) for \(n=3000\), \(5000\), and \(7000\), respectively. The \(\theta_2\)-only estimator is also substantially less biased than the plug-in, TMLE, one-step, and ridge Riesz benchmarks, and it attains the smallest MSE among the collaborative estimators at \(n=7000\). The library estimator remains less biased than the non-collaborative benchmarks, but it does not uniformly select the most bias-reducing or most stable candidate, illustrating a finite-sample bias--stability trade-off.

Overall, the results should not be read as showing uniform dominance of CR-TMLE over all benchmarks. Rather, they support the more specific claim targeted by this experiment: in a setting with misspecified outcome regressions and many propensity-relevant but target-irrelevant instruments, collaborative representation learning can substantially reduce bias relative to plug-in, one-step, TMLE, and direct Riesz alternatives. The price is that the bias-reducing collaborative fits can have larger MSE, reflecting a genuine bias--variance trade-off. These findings are consistent with the motivation for a targeted collaborative selector: nuisance learning should balance estimated drift and variability for the parameter of interest, rather than optimize for prediction.

\section{Conclusion}

We introduced a target-aware alternative to stepwise CTMLE for causal effects under outcome selection. The estimator learns low-dimensional treatment and selection representations from drift-oriented pseudo-outcomes, uses a finite collaborative library selected by targeted cross-validation, and performs final targeting on held-out folds to preserve the empirical EIF equation. Methodologically, the proposal replaces order-dependent covariate search with representation learning that is explicitly tied to the bias of the target parameter. The theory clarifies when collaboration, rather than full propensity consistency, is sufficient for valid estimation, and it sharpens the exposure-side robustness condition by showing that arm-specific drift information is needed. The simulations illustrate the intended behavior in a challenging design with misspecified outcome regressions and many target-irrelevant instruments: collaborative representations can substantially reduce bias relative to plug-in, one-step, TMLE, and Riesz alternatives, although the most aggressive bias reduction may increase variance. These results motivate richer collaborative libraries and more stable selectors in future work. Overall, the paper shows that propensity learning for targeted causal inference should be organized around the estimand, not around nuisance prediction alone.



%
%
%
\bibliographystyle{splncsNoDOI}
\bibliography{mybibliography}

@article{HuberDW,
  title = {Identification of Average Treatment Effects in Social Experiments Under Alternative Forms of Attrition},
  author = {Martin Huber},
  year = 2012,
  journal = {Journal of Educational and Behavioral Statistics},
  volume = 37,
  number = 3,
  pages = {443--474},
  doi = {10.3102/1076998611411917}
}

@inproceedings{Mohan2013,
  title = {Graphical Models for Inference with Missing Data},
  author = {Karthika Mohan and Judea Pearl and Jin Tian},
  year = 2013,
  booktitle = {Advances in Neural Information Processing Systems},
  publisher = {Curran Associates, Inc.},
  volume = 26,
  series = {{NeurIPS}},
  editor = {C. Burges and L. Bottou and M. Welling and Z. Ghahramani and K. Weinberger}
}

@inproceedings{Mohan2014,
  title = {Graphical Models for Recovering Probabilistic and Causal Queries from Missing Data},
  author = {Karthika Mohan and Judea Pearl},
  year = 2014,
  booktitle = {Advances in Neural Information Processing Systems},
  publisher = {Curran Associates, Inc.},
  volume = 27,
  editor = {Z. Ghahramani and M. Welling and C. Cortes and N. Lawrence and K. Weinberger},
  series = {{NeurIPS}}
}

@inproceedings{saadati2019adjustment,
  title = {Adjustment criteria for recovering causal effects from missing data},
  author = {Mojdeh Saadati and Jin Tian},
  year = 2019,
  booktitle = {Joint European Conference on Machine Learning and Knowledge Discovery in Databases},
  pages = {561--577},
  publisher = {Springer},
  series = {{ECML PKDD}},
  doi = {10.1007/978-3-030-46150-8_33}
}

@inproceedings{Bhattacharya2020,
  title = {Identification in Missing Data Models Represented by Directed Acyclic Graphs},
  author = {Rohit Bhattacharya and Razieh Nabi and Ilya Shpitser and James Robins},
  year = 2020,
  booktitle = {Proceedings of the 35th Conference on Uncertainty in Artificial Intelligence},
  series = {{UAI}},
  publisher = {{PMLR}},
  volume = 115,
  pages = {1149--1158},
  editor = {Ryan Adams and Vibhav Gogate}
}

@inproceedings{NabiFull,
  title = {Full law identification in graphical models of missing data: completeness results},
  author = {Razieh Nabi and Rohit Bhattacharya and Ilya Shpitser},
  year = 2020,
  booktitle = {Proceedings of the 37th International Conference on Machine Learning},
  articleno = 663,
  series = {{ICML}},
  publisher = {{PMLR}},
  pages = {7153--7163}
}

@inproceedings{NabiSemi,
  title = {Sufficient identification conditions and semiparametric estimation under missing not at random mechanisms},
  author = {Anna Guo and Jiwei Zhao and Razieh Nabi},
  booktitle = {Proceedings of the Thirty-Ninth Conference on Uncertainty in Artificial Intelligence},
  pages = {777--787},
  year = 2023,
  editor = {Robin Evans and Ilya Shpitser},
  volume = 216,
  publisher = {{PMLR}},
  series = {{UAI}}
}

@article{mgraphs,
  title = {Graphical Models for Processing Missing Data},
  author = {Karthika Mohan and Judea Pearl},
  year = 2021,
  journal = {Journal of the American Statistical Association},
  publisher = {Taylor and Francis},
  volume = 116,
  number = 534,
  pages = {1023--1037},
  doi = {10.1080/01621459.2021.1874961}
}

@article{Huber,
  title = {Double Machine Learning for Sample Selection Models},
  author = {Michela Bia and Martin Huber and Luk{\'a}{\v s} Laff{\'e}rs},
  year = 2024,
  journal = {Journal of Business and Economic Statistics},
  volume = 42,
  number = 3,
  pages = {958--969},
  doi = {10.1080/07350015.2023.2271071}
}

@inproceedings{Bareinboim12,
  title = {Controlling Selection Bias in Causal Inference},
  author = {Elias Bareinboim and Judea Pearl},
  year = 2012,
  booktitle = {Proceedings of the Fifteenth International Conference on Artificial Intelligence and Statistics},
  address = {La Palma, Canary Islands},
  publisher = {{PMLR}},
  series = {{AISTATS}},
  volume = 22,
  pages = {100--108},
  editor = {Neil Lawrence and Mark Girolami}
}

@article{Correa_Tian_Bareinboim_2018,
  title = {Generalized Adjustment Under Confounding and Selection Biases},
  author = {Juan Correa and Jin Tian and Elias Bareinboim},
  year = 2018,
  journal = {Proceedings of the {{AAAI}} Conference on Artificial Intelligence},
  volume = 32,
  number = 1,
  doi = {10.1609/aaai.v32i1.12125}
}

@article{tikka2021causal,
  title = {Causal Effect Identification from Multiple Incomplete Data Sources: A General Search-Based Approach},
  author = {Santtu Tikka and Antti Hyttinen and Juha Karvanen},
  year = 2021,
  journal = {Journal of Statistical Software},
  volume = 99,
  pages = {1--40},
  doi = {10.18637/jss.v099.i05}
}

@article{MLDR,
  title = {Machine learning in the estimation of causal effects: targeted minimum loss-based estimation and double/debiased machine learning},
  author = {Iv{\'a}n D{\'\i}az},
  year = 2019,
  journal = {Biostatistics},
  volume = 21,
  number = 2,
  pages = {353--358},
  doi = {10.1093/biostatistics/kxz042},
  issn = {1465-4644}
}

@article{AIPW2017,
  title = {Inverse-Probability-Weighted Estimation for Monotone and Nonmonotone Missing Data},
  author = {BaoLuo Sun and Neil Perkins and Stephen Cole and Ofer Harel and Emily Mitchell and Enrique Schisterman and Eric {Tchetgen Tchetgen}},
  year = 2017,
  journal = {American Journal of Epidemiology},
  volume = 187,
  number = 3,
  pages = {585--591},
  doi = {10.1093/aje/kwx350},
  issn = {0002-9262}
}

@article{biele2019bias,
  title = {Bias from self selection and loss to follow-up in prospective cohort studies},
  author = {Guido Biele and Kristin Gustavson and Nikolai Czajkowski and Roy Nilsen and Ted Reichborn and Per Magnus and Camilla Stoltenberg and Heidi Aase},
  year = 2019,
  journal = {European Journal of Epidemiology},
  publisher = {Springer},
  volume = 34,
  number = 10,
  pages = {927--938},
  doi = {10.1007/s10654-019-00550-1}
}

@article{ours,
  doi = {10.1515/jci-2024-0009},
  title = {Recovery and inference of causal effects with sequential adjustment for confounding and attrition},
  author = {Johan {de Aguas} and Johan Pensar and Tomás {Varnet Pérez} and Guido Biele},
  pages = 20240009,
  volume = 13,
  number = 1,
  journal = {Journal of Causal Inference},
  year = 2025
}

@article{Wei2022,
  title = {Doubly robust estimation in causal inference with missing outcomes: With an application to the Aerobics Center Longitudinal Study},
  author = {Kecheng Wei and Guoyou Qin and Jiajia Zhang and Xuemei Sui},
  year = 2022,
  journal = {Computational Statistics \& Data Analysis},
  volume = 168,
  pages = 107399,
  doi = {10.1016/j.csda.2021.107399},
  issn = {0167-9473}
}

@article{mcisaac2017,
  title = {Statistical methods for incomplete data: results on model misspecification},
  author = {Michelle McIsaac and Richard Cook},
  year = 2017,
  journal = {Statistical Methods in Medical Research},
  volume = 26,
  number = 1,
  pages = {248--267},
  doi = {10.1177/0962280214544251},
  pmid = 25063681
}

@article{multiplyDR,
  title = {Multiply robust estimation of the average treatment effect with missing outcomes},
  author = {Kecheng Wei and Guoyou Qin and Jiajia Zhang and Xuemei Sui},
  year = 2023,
  journal = {Journal of Statistical Computation and Simulation},
  publisher = {Taylor \& Francis},
  volume = 93,
  number = 10,
  pages = {1479--1495},
  doi = {10.1080/00949655.2022.2143501}
}

@article{vanderLaanRubin2006,
  title = {Targeted Maximum Likelihood Learning},
  author = {Mark {van der Laan} and Daniel Rubin},
  year = 2006,
  journal = {The International Journal of Biostatistics},
  volume = 2,
  number = 1,
  doi = {10.2202/1557-4679.1043}
}

@article{gruber2009targeted,
  title = {Targeted Maximum Likelihood Estimation: A Gentle Introduction},
  author = {Susan Gruber and Mark {van der Laan}},
  year = 2009,
  journal = {The American Statistician},
  publisher = {Taylor and Francis},
  volume = 63,
  number = 4,
  pages = {1--38}
}

@article{attrition,
  title = {Attrition Bias Related to Missing Outcome Data: A Longitudinal Simulation Study},
  author = {Antoine Lewin and Ruben Brondeel and Tarik Benmarhnia and Fr{\'e}d{\'e}rique Thomas and Basile Chaix},
  year = 2018,
  journal = {Epidemiology},
  volume = 29,
  number = 1,
  doi = {10.1097/ede.0000000000000755}
}

@article{dw,
  title = {Doubly weighted M-estimation for nonrandom assignment and missing outcomes},
  author = {Akanksha Negi},
  year = 2024,
  journal = {Journal of Causal Inference},
  volume = 12,
  number = 1,
  pages = 20230016,
  doi = {10.1515/jci-2023-0016}
}

@article{DML,
  title = {Double/debiased machine learning for treatment and structural parameters},
  author = {Victor Chernozhukov and Denis Chetverikov and Mert Demirer and Esther Duflo and Christian Hansen and Whitney Newey and James Robins},
  year = 2018,
  journal = {The Econometrics Journal},
  volume = 21,
  number = 1,
  pages = {C1--C68},
  doi = {10.1111/ectj.12097},
  issn = {1368-4221}
}

@article{schnitzer2016variable,
  title = {Variable selection for confounder control, flexible modeling and collaborative targeted minimum loss-based estimation in causal inference},
  author = {Mireille Schnitzer and Judith Lok and Susan Gruber},
  year = 2016,
  journal = {The International Journal of Biostatistics},
  publisher = {De Gruyter},
  volume = 12,
  number = 1,
  pages = {97--115},
  doi = {10.1515/ijb-2015-0017}
}

@article{nuisanceModels,
  title = {The Costs and Benefits of Uniformly Valid Causal Inference with High-Dimensional Nuisance Parameters},
  author = {Niloofar Moosavi and Jenny H{\"a}ggstr{\"o}m and Xavier {de Luna}},
  year = 2023,
  journal = {Statistical Science},
  publisher = {Institute of Mathematical Statistics},
  volume = 38,
  number = 1,
  pages = {1--12},
  doi = {10.1214/21-sts843}
}

@article{instability,
  title = {Comment: Performance of Double-Robust Estimators When Inverse Probability Weights Are Highly Variable},
  author = {James Robins and Mariela Sued and Quanhong Lei-Gomez and Andrea Rotnitzky},
  year = 2007,
  journal = {Statistical Science},
  publisher = {Institute of Mathematical Statistics},
  volume = 22,
  number = 4,
  pages = {544--559},
  doi = {10.1214/07-sts227d}
}

@article{seaman2013review,
  title = {Review of inverse probability weighting for dealing with missing data},
  author = {Shaun Seaman and Ian White},
  year = 2013,
  journal = {Statistical Methods in Medical Research},
  volume = 22,
  number = 3,
  pages = {278--295},
  doi = {10.1177/0962280210395740}
}

@article{superlearner2007,
  title = {Super Learner},
  author = {Mark {van der Laan} and Eric Polley and Alan Hubbard},
  year = 2007,
  journal = {Statistical Applications in Genetics and Molecular Biology},
  volume = 6,
  number = 1,
  doi = {10.2202/1544-6115.1309}
}

@article{stableIPW,
  title = {Stable inverse probability weighting estimation for longitudinal studies},
  author = {Vahe Avagyan and Stijn Vansteelandt},
  year = 2021,
  journal = {Scandinavian Journal of Statistics},
  volume = 48,
  number = 3,
  pages = {1046--1067},
  doi = {10.1111/sjos.12542}
}

@article{collaborative2010,
  title = {Collaborative double robust targeted maximum likelihood estimation},
  author = {Mark {van der Laan} and Susan Gruber},
  year = 2010,
  journal = {The International Journal of Biostatistics},
  publisher = {De Gruyter},
  volume = 6,
  number = 1,
  doi = {10.2202/1557-4679.1181}
}

@inproceedings{chernozhukov2022riesznet,
  title = {{RieszNet} and {ForestRiesz}: Automatic Debiased Machine Learning with Neural Nets and Random Forests},
  author = {Victor Chernozhukov and Whitney Newey and V\'{\i}ctor {Quintas-Mart\'{\i}nez} and Vasilis Syrgkanis},
  booktitle = {Proceedings of the 39th International Conference on Machine Learning},
  pages = {3901--3914},
  year = 2022,
  editor = {Kamalika Chaudhuri and Stefanie Jegelka and Le Song and Csaba Szepesvari and Gang Niu and Sivan Sabato},
  volume = 162,
  publisher = {{PMLR}},
  series = {{ICML}}
}

@inproceedings{Lars1,
  title = {Causal Isotonic Calibration for Heterogeneous Treatment Effects},
  author = {Lars {van der Laan} and Ernesto {Ulloa-Pérez} and Marco Carone and Alex Luedtke},
  booktitle = {Proceedings of the 40th International Conference on Machine Learning},
  pages = {34831--34854},
  year = 2023,
  editor = {Andreas Krause and Emma Brunskill and Kyunghyun Cho and Barbara Engelhardt and Sivan Sabato and Jonathan Scarlett},
  volume = 202,
  publisher = {{PMLR}},
  series = {{ICML}}
}

@inproceedings{Lars3,
  title = {Stabilized Inverse Probability Weighting via Isotonic Calibration},
  author = {Lars {van der Laan} and Ziming Lin and Marco Carone and Alex Luedtke},
  booktitle = {Proceedings of the Fourth Conference on Causal Learning and Reasoning},
  pages = {139--173},
  year = 2025,
  editor = {Biwei Huang and Mathias Drton},
  volume = 275,
  publisher = {{PMLR}},
  series = {{CLeaR}}
}

@article{scallable,
  title = {Scalable collaborative targeted learning for high-dimensional data},
  author = {Cheng Ju and Susan Gruber and Samuel Lendle and Antoine Chambaz and Jessica Franklin and Richard Wyss and Sebastian Schneeweiss and Mark {van der Laan}},
  year = 2019,
  journal = {Statistical Methods in Medical Research},
  volume = 28,
  number = 2,
  pages = {532--554},
  doi = {10.1177/0962280217729845}
}

@article{kato2026riesz,
  title = {{Riesz} representer fitting under {Bregman} divergence: A unified framework for debiased machine learning},
  author = {Masahiro Kato},
  journal = {{arXiv} preprint {arXiv}:2601.07752},
  year = 2026,
  doi = {10.48550/arXiv.2601.07752}
}

@article{chernozhukov2022debiased,
  title = {Debiased machine learning of global and local parameters using regularized {Riesz} representers},
  author = {Victor Chernozhukov and Whitney Newey and Rahul Singh},
  journal = {The Econometrics Journal},
  volume = 25,
  number = 3,
  pages = {576--601},
  year = 2022,
  publisher = {Oxford University Press},
  doi = {10.1093/ectj/utac002}
}

@inproceedings{chasm,
  author = {Benjamin Rhodes and Kai Xu and Michael Gutmann},
  title = {Telescoping density-ratio estimation},
  year = 2020,
  isbn = 9781713829546,
  publisher = {Curran Associates Inc.},
  address = {Red Hook, NY, USA},
  booktitle = {Proceedings of the 34th International Conference on Neural Information Processing Systems},
  articleno = 412,
  numpages = 12,
  location = {Vancouver, BC, Canada},
  series = {{NeurIPS}}
}

@inproceedings{KatoHack,
  title = {Non-Negative {Bregman} Divergence Minimization for Deep Direct Density Ratio Estimation},
  author = {Masahiro Kato and Takeshi Teshima},
  booktitle = {Proceedings of the 38th International Conference on Machine Learning},
  pages = {5320--5333},
  year = 2021,
  editor = {Marina Meila and Tong Zhang},
  volume = 139,
  series = {{ICML}},
  publisher = {{PMLR}}
}

@book{PearlCausality,
  title = {Causality},
  author = {Judea Pearl},
  year = 2009,
  publisher = {Cambridge University Press},
  address = {Cambridge, UK},
  doi = {10.1017/cbo9780511803161},
  isbn = {978-0-521-89560-6},
  edition = 2
}

@book{targetedlearning2011,
  title = {Targeted Learning: Causal Inference for Observational and Experimental Data},
  author = {Mark {van der Laan} and Sherri Rose},
  year = 2011,
  publisher = {Springer},
  address = {New York},
  series = {Springer Series in Statistics},
  doi = {10.1007/978-1-4419-9782-1}
}

@incollection{cvtmle2011,
  title = {Cross-Validated Targeted Minimum-Loss-Based Estimation},
  author = {Wenjing Zheng and Mark {van der Laan}},
  booktitle = {Targeted Learning: Causal Inference for Observational and Experimental Data},
  editor = {Mark {van der Laan} and Sherri Rose},
  pages = {459--474},
  year = 2011,
  publisher = {Springer},
  address = {New York, NY},
  series = {Springer Series in Statistics},
  doi = {10.1007/978-1-4419-9782-1_27}
}

@article{vdv2006oracle,
  title = {Oracle Inequalities for Multi-Fold Cross Validation},
  author = {Aad {van der Vaart} and Sandrine Dudoit and Mark {van der Laan}},
  year = 2006,
  journal = {Statistics \& Decisions},
  volume = 24,
  number = 3,
  pages = {351--371},
  doi = {10.1524/stnd.2006.24.3.351}
}

@inproceedings{shalit2017,
  title = {Estimating Individual Treatment Effect: Generalization Bounds and Algorithms},
  author = {Uri Shalit and Fredrik Johansson and David Sontag},
  year = 2017,
  booktitle = {Proceedings of the 34th International Conference on Machine Learning},
  series = {Proceedings of Machine Learning Research},
  volume = 70,
  pages = {3076--3085},
  publisher = {{PMLR}}
}

@article{johansson2022,
  title = {Generalization Bounds and Representation Learning for Estimation of Potential Outcomes and Causal Effects},
  author = {Fredrik Johansson and Uri Shalit and Nathan Kallus and David Sontag},
  year = 2022,
  journal = {Journal of Machine Learning Research},
  volume = 23,
  number = 166,
  pages = {1--50}
}

@article{shortreed2017oal,
  title = {Outcome-Adaptive Lasso: Variable Selection for Causal Inference},
  author = {Susan Shortreed and Ashkan Ertefaie},
  year = 2017,
  journal = {Biometrics},
  volume = 73,
  number = 4,
  pages = {1111--1122},
  doi = {10.1111/biom.12679}
}

@article{luo2017sdr,
  title = {On Estimating Regression-Based Causal Effects Using Sufficient Dimension Reduction},
  author = {Wei Luo and Yeying Zhu and Debashis Ghosh},
  year = 2017,
  journal = {Biometrika},
  volume = 104,
  number = 1,
  pages = {51--65},
  doi = {10.1093/biomet/asw068}
}
%

\clearpage


\appendix

\title{Collaborative Representations for Targeted Causal Inference under Outcome Selection}
\subtitle{Supplementary Material}

\titlerunning{Collaborative Representations for Targeted Causal Inference}

\author{}

\authorrunning{}

\institute{}

\maketitle

\section{Additional details and proofs}

\subsection{Proof of collaborative robustness}

Let $\widetilde\eta=(\widetilde Q_1,\widetilde Q_2,\widetilde\pi_A,\widetilde\pi_R)$ denote fixed putative nuisance functions, and let $\psi^\circ=\Psi[P^\circ]$. Write
\[
\delta_a(W):=\widetilde Q_2(W,a)-Q_2^\circ(W,a),\qquad a\in\{0,1\},
\]
and
\[
\varepsilon(X):=\widetilde Q_1(X)-Q_1^\circ(X),\qquad X=(W,A,Z).
\]
The efficient-influence-function form in the main text, $D(\widetilde\eta,\psi^\circ;O)$ is
\[
\frac{R\widetilde\alpha_2(W,A)}{\widetilde\pi_R(X)}\{Y^\dagger-\widetilde Q_1(X)\}
+
\widetilde\alpha_2(W,A)\{\widetilde Q_1(X)-\widetilde Q_2(W,A)\}
+
\Delta_a\widetilde Q_2(W,a)-\psi^\circ,
\]
where
\[
\widetilde\alpha_2(W,A)=
\frac{A-\widetilde\pi_A(W)}{\widetilde\pi_A(W)\{1-\widetilde\pi_A(W)\}}.
\]
Using $\E^\circ
[Y^\dagger-Q_1^\circ(X)\mid X,R=1]=0$ and $Q_2^\circ(W,A)=\E^\circ\{Q_1^\circ(X)\mid W,A\}$, the drift of the EIF estimating equation is
\begin{align}
\operatorname{drift}(\widetilde\eta,\eta^\circ)
&:=
\E^\circ D(\widetilde\eta,\psi^\circ;O)
\nonumber\\
&=
\E^\circ\!\left[
\frac{A-\widetilde\pi_A(W)}{\widetilde\pi_A(W)\{1-\widetilde\pi_A(W)\}}
\left(1-\frac{R}{\widetilde\pi_R(X)}\right)
\varepsilon(X)
\right]
\nonumber\\
&\quad+
\E^\circ\!\left[
\frac{\widetilde\pi_A(W)-A}{\widetilde\pi_A(W)}\,\delta_1(W)
\right]
+
\E^\circ\!\left[
\frac{\widetilde\pi_A(W)-A}{1-\widetilde\pi_A(W)}\,\delta_0(W)
\right].
\label{eq:app-drift-decomp}
\end{align}
This is the decomposition used by the collaborative robustness theorem.

For the exposure-side terms, suppose
\[
\widetilde\pi_A(W)=\pr^\circ(A=1\mid \delta_1(W),\delta_0(W)).
\]
Then $\widetilde\pi_A(W)$ is measurable with respect to the sigma-field generated by $\delta_1(W)$ and $\delta_0(W)$, and
\[
\E^\circ[A-\widetilde\pi_A(W)\mid \delta_1(W),\delta_0(W)]=0.
\]
The two exposure-side expectations in \eqref{eq:app-drift-decomp} are therefore zero, because their multipliers are measurable functions of $(\delta_1(W),\delta_0(W))$.

For the selection-side term, condition first on $(W,A)$:
\begin{align*}
&\E^\circ\!\left[
\frac{A-\widetilde\pi_A(W)}{\widetilde\pi_A(W)\{1-\widetilde\pi_A(W)\}}
\left(1-\frac{R}{\widetilde\pi_R(X)}\right)
\varepsilon(X)
\right]
\\
&\quad=
\E^\circ\!\left[
\frac{A-\widetilde\pi_A(W)}{\widetilde\pi_A(W)\{1-\widetilde\pi_A(W)\}}
\E^\circ\!\left\{\left(1-\frac{R}{\widetilde\pi_R(X)}\right)\varepsilon(X)\mid W,A\right\}
\right].
\end{align*}
If
\[
\widetilde\pi_R(X)=\pr^\circ (R=1\mid W,A,\varepsilon(X)),
\]
then
\[
\E^\circ\!\left\{\frac{R}{\widetilde\pi_R(X)}\varepsilon(X)\mid W,A,\varepsilon(X)\right\}=\varepsilon(X),
\]
and hence the inner conditional expectation given $(W,A)$ is zero. Thus all terms in \eqref{eq:app-drift-decomp} vanish and
\[
\E^\circ D(\widetilde\eta,\psi^\circ;O)=0.
\]
This proves the stated sufficient collaborative robustness condition.

\begin{remark}[Conditioning on the pair, not only on the contrast]\label{rem:app-pair-not-contrast}
The exposure-side condition cannot generally be weakened to conditioning on the scalar contrast $\delta_1(W)-\delta_0(W)$. The two arm-specific drifts enter \eqref{eq:app-drift-decomp} through different inverse-probability multipliers, so balancing $A$ against the contrast alone does not force both conditional expectations to vanish. The scalar contrast is sufficient only under an additional restriction making $(\delta_1,\delta_0)$ measurable with respect to $\delta_1-\delta_0$, for example a common-drift restriction tying the two arms deterministically.
\end{remark}

\begin{remark}[Why the attrition condition keeps $(W,A)$]\label{rem:app-attrition-conditioning}
The attrition term is multiplied by the exposure-side clever covariate $\alpha_2(W,A)$, so the relevant inner expectation is over $(Z,R)$ conditionally on $(W,A)$. A specification of the form $\widetilde\pi_R=\pr^\circ (R=1\mid \varepsilon(X))$ is therefore insufficient unless $\pr^\circ (R=1\mid W,A,\varepsilon(X))$ is already a function of $\varepsilon(X)$ alone. The skip connection for $(W,A)$ in the attrition head is included for this reason.
\end{remark}

\subsection{Verification of the collaborative pseudo-outcomes}\label{app:pseudo}

The pseudo-outcomes in the main text are designed to provide observable noisy proxies for the drift functions in the preceding theorem. Throughout this subsection, the working nuisances are treated as fixed and are understood to be cross-fitted in implementation.

For the selection component, define among complete cases
\[
e_R=\widetilde Q_1(X)-Y^\dagger.
\]
Then
\[
\E^\circ[e_R\mid X,R=1]
=
\widetilde Q_1(X)-\E^\circ[Y^\dagger\mid X,R=1]
=
\widetilde Q_1(X)-Q_1^\circ(X)
=
\varepsilon(X).
\]
Thus $e_R$ is a direct complete-case proxy for the first-stage drift. The corresponding auxiliary loss recovers the projection of $\varepsilon(X)$ under the complete-case law. To use the learned representation for the full attrition propensity, one still needs selection positivity and enough regularity or extrapolation so that the drift direction learned among $R=1$ units is informative on the validation support. This is an empirical limitation of any proxy based on observed outcomes, not an additional identification formula.

For the exposure component, define
\[
\xi
=
\widetilde Q_1(X)
+
\frac{R}{\widetilde\pi_R(X)}\{Y^\dagger-\widetilde Q_1(X)\}.
\]
Conditional on $X$, this has mean
\[
\E^\circ[\xi\mid X]
=
Q_1^\circ(X)+
\left\{1-\frac{\pi_R^\circ(X)}{\widetilde\pi_R(X)}\right\}
\varepsilon(X).
\]
Consequently,
\[
\E^\circ[\xi\mid W,A]
=
Q_2^\circ(W,A)
+
\E^\circ\!\left[
\left\{1-\frac{\pi_R^\circ(X)}{\widetilde\pi_R(X)}\right\}\varepsilon(X)
\mid W,A\right].
\]
The second term is zero when $\widetilde\pi_R$ is the true selection propensity, and it is also zero under the selection-side collaborative condition proved above. In either case,
\[
\E^\circ[\xi\mid W,A=a]=Q_2^\circ(W,a).
\]
With
\[
e_A=\widetilde Q_2(W,A)-\xi,
\]
it follows that
\[
\E^\circ[e_A\mid W,A=a]=\widetilde Q_2(W,a)-Q_2^\circ(W,a)=\delta_a(W),
\qquad a\in\{0,1\}.
\]
This verifies that the exposure-side auxiliary regression should be vector-valued, with separate heads for the two factual arms. If the preceding selection correction is only approximate, then $e_A$ targets $\delta_a$ plus the residual selection-correction drift; the targeted validation criterion is included precisely to select against candidates for which this residual drift remains large.

\begin{remark}[Sequential nature of the exposure proxy]\label{rem:app-eA-sequential}
The proxy $e_A$ is arm-specific only after the preceding selection correction is adequate, either by a correct selection model or by the selection-side collaborative condition. Otherwise it carries residual first-stage selection drift as well as second-stage treatment drift.
\end{remark}

\subsection{Population interpretation of the collaborative penalties}\label{app:penalties}

For a representation $u_R=h_R(X)$, the population version of the selection penalty is
\[
\E^\circ\!\left[R\{e_R-r(u_R)\}^2\right].
\]
Its unrestricted minimizer is
\[
r_R^\star(u)=\E^\circ[e_R\mid h_R(X)=u,R=1].
\]
Thus, if $h_R$ is sufficiently rich to retain the drift direction, $r_R^\star\{h_R(X)\}$ equals the complete-case projection of $\varepsilon(X)$. The attrition propensity head then receives $(W,A,h_R(X))$, matching the sufficient form $\pr^\circ (R=1\mid W,A,\varepsilon(X))$ up to representation approximation.

For the exposure component, let
\[
\widetilde\pi_A(1\mid W)=\widetilde\pi_A(W),
\qquad
\widetilde\pi_A(0\mid W)=1-\widetilde\pi_A(W).
\]
For each arm $a$, the population version of the weighted auxiliary loss is
\[
\E^\circ\!\left[
\frac{\mathbb{I}(A=a)}{\widetilde\pi_A(a\mid W)}
\{e_A-r_a(h_A(W))\}^2
\right].
\]
Its pointwise minimizer is the weighted projection
\[
r_{A,a}^\star(u)=
\frac{
\E^\circ[\mathbb I(A=a)\widetilde\pi_A(a\mid W)^{-1}e_A\mid h_A(W)=u]
}{
\E^\circ[\mathbb I(A=a)\widetilde\pi_A(a\mid W)^{-1}\mid h_A(W)=u]
}.
\]
When the selection correction is adequate and the working treatment propensity is close to the true one, this approximates a marginal-in-$W$ projection of the arm-specific drifts.

The bottleneck penalties in the main text constrain the encoders rather than the propensity heads. Because neural encoders and heads are scale-nonidentifiable, a practical implementation should standardize latent codes or use coordinated weight decay; otherwise an encoder penalty can be offset by rescaling the downstream head without changing the propensity score.

\subsection{Relation to Riesz regression}\label{app:riesz}

The collaborative auxiliary losses are related to Riesz regression, but they solve a different problem. To make the distinction explicit, write the residual part of the recovered-ATE influence function as
\[
\alpha_1(O)\{Y^\dagger-Q_1(X)\}
+
\alpha_2(O)\{Q_1(X)-Q_2(W,A)\}.
\]
A direct Riesz-regression approach starts from a linear functional of regression errors and estimates a representer for that functional. For example, for a function class \(\mathcal F_1\) for first-stage residuals and \(\mathcal F_2\) for second-stage residuals, the ideal representers \(H_1^\star\) and \(\alpha_2^\star\), with \(H_1^\circ(X,A)=\alpha_2^\circ(W,A)/\pi_R^\circ(X)\), satisfy the normal equations
\[
\E^\circ\{R f_1(X)H_1^\star(X,A)\}
=
\E^\circ\{R f_1(X)H_1^\circ(X,A)\},
\qquad f_1\in\mathcal F_1,
\]
\[
\E^\circ\{f_2(W,A)\alpha_2^\star(O)\}
=
\E^\circ\{f_2(W,A)\alpha_2^\circ(O)\},
\qquad f_2\in\mathcal F_2.
\]
Equivalently, if \(\Phi_1(X,A)\) and \(\Phi_2(O)\) are finite bases, the ridge version used in the numerical benchmark estimates
\[
\widehat\beta_1
=
\arg\min_\beta
\E_n\{R[H_1(\widehat\pi)-\Phi_1^\top\beta]^2\}
+\lambda\|\beta\|_2^2,
\qquad
\widehat H_1^\star(X,A)=\Phi_1(X,A)^\top\widehat\beta_1,
\]
\[
\widehat\beta_2
=
\arg\min_\beta
\E_n\{\alpha_2(\widehat\pi)-\Phi_2^\top\beta\}^2
+\lambda\|\beta\|_2^2,
\qquad
\widehat\alpha_2^\star(O)=\Phi_2(O)^\top\widehat\beta_2.
\]
The resulting estimator replaces the inverse-probability weights in the one-step correction by these learned representers,
\[
\E_n\left[
\widehat Q_2(W,1)-\widehat Q_2(W,0)
+R\widehat H_1^\star(X,A)\{Y^\dagger-\widehat Q_1(X)\}
+\widehat\alpha_2^\star(O)\{\widehat Q_1(X)-\widehat Q_2(W,A)\}
\right].
\]
Thus direct Riesz regression is a \emph{dual-side} procedure: it learns weights that represent the linearized target functional.

The proposed collaborative representation learner instead uses Riesz-like square-loss projections on the \emph{drift side}. The attrition auxiliary head solves, at the population level,
\[
r_R^\star
=
\arg\min_r
\E^\circ\!\left[R\{e_R-r(h_R(X))\}^2\right],
\quad
r_R^\star\{h_R(X)\}
=
\E^\circ[e_R\mid h_R(X),R=1].
\]
Because \(\E^\circ[e_R\mid X,R=1]=\varepsilon(X)\), this is a projection of the first-stage drift onto the sigma-field generated by the attrition representation. Similarly, for the exposure component,
\[
r_{A,a}^\star
=
\arg\min_r
\E^\circ\!\left[
\frac{\mathbb{I}(A=a)}{\widetilde\pi_A(a\mid W)}
\{e_A-r(h_A(W))\}^2
\right],
\]
and its pointwise solution is the weighted conditional projection
\[
r_{A,a}^\star\{h_A(W)\}
=
\frac{
\E^\circ\!\left[\mathbb{I}(A=a)\widetilde\pi_A(a\mid W)^{-1}e_A\mid h_A(W)\right]
}{
\E^\circ\!\left[\mathbb{I}(A=a)\widetilde\pi_A(a\mid W)^{-1}\mid h_A(W)\right]
}.
\]
When the selection correction is adequate and \(\widetilde\pi_A\) is the true treatment propensity, this reduces to the marginal projection \(\E^\circ\{\delta_a(W)\mid h_A(W)\}\). For general working propensities, it is a weighted projection of the same arm-specific drift. These auxiliary projections are not inserted as final Riesz weights. They are used to constrain the representation available to the treatment and attrition propensity heads, so that \(\widehat\pi_A\) and \(\widehat\pi_R\) are learned along directions that are relevant to the target remainder.

This difference explains the methodological role of the collaborative penalties. Direct Riesz regression regularizes the correction term itself by approximating \((H_1,\alpha_2)\). The collaborative learner regularizes the nuisance scores by learning representations of \((\varepsilon,\delta_1,\delta_0)\), then forms the final estimator through the usual targeted substitution step and its associated influence-function equation. Consequently, the proposed estimator is not a Riesz estimator, although its auxiliary losses use the same Hilbert-space projection logic that underlies Riesz representation methods.

\subsection{Targeted selection criterion}\label{app:targeted-criterion}

For a candidate $s\in\mathcal S_n$ trained on $\mathcal T\setminus\mathcal T_v$ and evaluated on a disjoint inner validation fold $\mathcal T_v$, define
\[
\widehat\psi^s_{-v,\mathcal T_v}
=
\E_{\mathcal T_v}[\widehat Q^s_{2,-v}(W,1)-\widehat Q^s_{2,-v}(W,0)],
\]
\[
\widehat b_v(s)
=
\E_{\mathcal T_v}D(\widehat\eta^s_{-v},\widehat\psi^s_{-v,\mathcal T_v};O),
\qquad
\widehat v_v(s)
=
\E_{\mathcal T_v}D(\widehat\eta^s_{-v},\widehat\psi^s_{-v,\mathcal T_v};O)^2.
\]
Because the candidate is trained without using $\mathcal T_v$, the held-out mean is an honest estimate of the signed population drift, up to the plug-in centering error:
\[
\E\{\widehat b_v(s)\mid \mathcal T\setminus\mathcal T_v\}
=
\operatorname{drift}(\widehat\eta^s_{-v},\eta^\circ)+O_{P^\circ}(|\mathcal T_v|^{-1/2}).
\]
Consequently, when $\zeta=0$ and uniformly over a finite or slowly growing library,
\[
\mathcal R_n(s)
=
\operatorname{drift}(\widehat\eta^s,\eta^\circ)^2
+O_{P^\circ}(|\mathcal T|^{-1})
+o_{P^\circ}\{\operatorname{drift}(\widehat\eta^s,\eta^\circ)^2+|\mathcal T|^{-1}\},
\]
under the usual boundedness and maximal-inequality conditions. Thus the leading criterion controls squared drift with a first-order variance penalty. Since it controls the square of a signed drift, the main theorem states the oracle-transfer condition in terms of $|\operatorname{drift}|$. If the optional regression-loss component dominates the criterion, or if the selected candidate is refit after selection, local equivalence to the final target remainder is not automatic and must be imposed as a stability condition.

\begin{remark}[Squared held-out mean]\label{rem:app-squared-mean}
The term $\widehat b_n(s)^2$ is upward biased for squared drift: conditionally on the training folds, $\E\{\widehat b_n(s)^2\}=\operatorname{drift}^2+O(|\mathcal T|^{-1})$. Hence \eqref{eq:targeted-risk} mildly over-penalizes variance by an $O(|\mathcal T|^{-1})$ term, unless one subtracts an estimated variance for $\widehat b_n(s)$. This conservative bias vanishes and does not affect the oracle-transfer argument.
\end{remark}

\begin{remark}[Library growth]\label{rem:app-library-growth}
Selection over the library contributes the usual maximal-inequality factor of order $\sqrt{\log N_n}$. Fixed grids and many polynomially growing grids are therefore harmless for the first-order expansion under bounded or sub-Gaussian envelopes. Aggressive architecture search or adaptive early stopping outside the declared library should instead be treated as part of the learning algorithm and requires its own stability argument.
\end{remark}

\subsection{Proof of asymptotic validity}\label{app:asymptotics}

We write the proof using the pooled outer validation folds. For each observation $i$, let $k(i)$ denote its outer validation fold, and let $\widehat\eta_{k(i)}^{\widehat s,\sharp}$ be the nuisance tuple obtained by training and selecting on the corresponding outer training sample and then applying the final validation-fold fluctuation. The targeting step gives, fold by fold,
\[
\E_{\mathcal E_k}D(\widehat\eta_k^{\widehat s,\sharp},\widehat\psi_k^\sharp;O)=o_{P^\circ}(n^{-1/2})
\]
up to numerical optimization error. Since $D(\eta,\widehat\psi;O)=D(\eta,\psi^\circ;O)+\psi^\circ-\widehat\psi$, the cross-fitted estimator satisfies the algebraic expansion
\begin{equation}\label{eq:app-cvtmle-expansion}
\widehat\psi^\sharp-\psi^\circ
=
\frac{1}{n}\sum_{i=1}^n
D(\widehat\eta_{k(i)}^{\widehat s,\sharp},\psi^\circ;O_i)
+o_{P^\circ}(n^{-1/2}).
\end{equation}
Decompose the summand around the probability limit $\bar\eta$ of the selected validation-targeted nuisance tuple:
\[
D(\widehat\eta_{k(i)}^{\widehat s,\sharp},\psi^\circ;O_i)
=
D(\bar\eta,\psi^\circ;O_i)
+
F(\widehat\eta_{k(i)}^{\widehat s,\sharp},\bar\eta,\psi^\circ;O_i),
\]
where
\[
F(\widehat\eta,\bar\eta,\psi^\circ;O)=D(\widehat\eta,
\psi^\circ;O)-D(\bar\eta,
\psi^\circ;O).
\]
The empirical-process assumption (A5) gives
\[
\frac{1}{n}\sum_{i=1}^n
\left[
F(\widehat\eta_{k(i)}^{\widehat s,\sharp},\bar\eta,\psi^\circ;O_i)
-
\E^\circ F(\widehat\eta_{k(i)}^{\widehat s,\sharp},\bar\eta,\psi^\circ;O)
\right]
=o_{P^\circ}(n^{-1/2}).
\]
The remaining expectation is the selected remainder, up to continuity at $\bar\eta$:
\[
\E^\circ D(\widehat\eta^{\widehat s,\sharp},\psi^\circ;O)
=
\operatorname{drift}(\widehat\eta^{\widehat s,\sharp},\eta^\circ).
\]
By the oracle inequality and the transfer condition (A4),
\[
|\operatorname{drift}(\widehat\eta^{\widehat s,\sharp},\eta^\circ)|
=
|\operatorname{drift}(\widehat\eta^{s_n^\dagger,\sharp},\eta^\circ)|+o_{P^\circ}(1).
\]
Assumption (A3) therefore implies that the selected remainder is $o_{P^\circ}(1)$, and by continuity $\E^\circ D(\bar\eta,\psi^\circ;O)=0$. Equation \eqref{eq:app-cvtmle-expansion} then gives
\[
\widehat\psi^\sharp-\psi^\circ
=
\frac{1}{n}\sum_{i=1}^n D(\bar\eta,\psi^\circ;O_i)+o_{P^\circ}(1),
\]
which proves consistency.

If the oracle remainder and the selection-transfer condition hold at the $n^{-1/2}$ scale, then
\[
|\operatorname{drift}(\widehat\eta^{\widehat s,\sharp},\eta^\circ)|=o_{P^\circ}(n^{-1/2}),
\]
and the expansion strengthens to
\[
\sqrt n(\widehat\psi^\sharp-\psi^\circ)
=
\frac{1}{\sqrt n}\sum_{i=1}^n D(\bar\eta,\psi^\circ;O_i)+o_{P^\circ}(1).
\]
The central limit theorem yields
\[
\sqrt n(\widehat\psi^\sharp-\psi^\circ)
\rightsquigarrow
N\!\left(0,\Var_{P^\circ}\{D(\bar\eta,\psi^\circ;O)\}\right).
\]
If, in addition, $\bar\eta=\eta^\circ$ and the usual second-order product-rate conditions hold, then $D(\bar\eta,\psi^\circ;O)$ is the efficient influence function and the estimator attains the semiparametric efficiency bound.

\begin{remark}[Inference when the nuisance limit is collaborative but not true]\label{rem:app-inference}
Under collaborative robustness, the selected nuisance limit can satisfy the condition $\E^\circ D(\bar\eta,\psi^\circ;O)=0$ without equaling the true nuisance tuple. The resulting first-order influence function is then $D(\bar\eta,\psi^\circ;O)$, not the efficient influence function at $\eta^\circ$. Variance estimation should therefore use the EIF-form estimating function evaluated at the selected validation-targeted fit, and efficiency should only be claimed under full nuisance consistency.
\end{remark}

\begin{remark}[Scope of the high-level conditions]\label{rem:app-high-level}
The theorem assumes oracle-transfer and stability properties for the selected finite-library candidate. It does not provide a separate learning-theoretic proof that the neural representation training or numerical optimizer recovers the oracle representation; those requirements are summarized in (A3)--(A5).
\end{remark}

\subsection{Variance estimation under repeated sample splitting}\label{app:variance}

For one outer cross-fitting split, the natural variance estimate is
\[
\widehat{\Var}_{\mathrm{IF}}
=
\frac{1}{n}\widehat{\Var}_n
\left\{D(\widehat\eta^{\widehat s,\sharp},\widehat\psi^\sharp;O)\right\},
\]
where each observation is evaluated with the nuisance tuple from its validation fold. If the whole procedure is repeated over $M$ independent split assignments, with estimates $\widehat\psi_m$, within-split variances $\widehat{\Var}_{\mathrm{IF},m}$, and average $\bar\psi=M^{-1}\sum_{m=1}^M\widehat\psi_m$, define
\[
V_W=\frac{1}{M}\sum_{m=1}^M\widehat{\Var}_{\mathrm{IF},m},
\qquad
V_B=\frac{1}{M-1}\sum_{m=1}^M(\widehat\psi_m-\bar\psi)^2.
\]
For the averaged point estimate, $V_W+V_B/M$ estimates the within-split influence-function variance plus Monte Carlo variability from averaging over random splits. The more conservative $V_W+V_B$ treats split instability as part of the uncertainty attached to the reported procedure. Reporting both is useful in weak-overlap settings because a large gap indicates that the selected path or representation is unstable across folds.

\section{Simulation design and implementation details}\label{app:simulation}

The simulation evaluates the recovered average treatment effect under nonlinear treatment assignment, post-treatment outcome selection, deliberately misspecified sequential outcome regressions, neural propensity score estimation, and collaborative representation learning. The implementation reported in the paper is the script \texttt{simulation\_run.py}. The numerical run uses
\begin{verbatim}
python3 simulation_run.py \
  --n-rep 100 \
  --sample-sizes 3000 5000 7000 \
  --budget 3 \
  --zeta 0 \
  --outer-folds 4 \
  --inner-folds 2 \
  --epochs 20 \
  --inner-epochs 10 \
  --batch-size 1024 \
  --device cpu \
  --torch-threads 4 \
  --riesz-alpha 1.0 \
  --clip-eps 0.001 \
  --output-dir simulation_output
\end{verbatim}
Thus the Monte Carlo experiment uses \(100\) repetitions at sample sizes \(n\in\{3000,5000,7000\}\), four outer cross-fitting folds, two inner folds for candidate selection, budget \(B=3\), and \(\zeta=0\). Neural propensity networks use hidden width \(64\), AdamW learning rate \(10^{-3}\), weight decay \(10^{-4}\), and batch size \(1024\). Outer neural propensity refits are trained for \(20\) epochs, while inner-CV propensity fits are trained for \(10\) epochs. The run uses CPU execution with four PyTorch threads. All inverse-probability terms use propensity clipping at \(\epsilon_{\rm clip}=0.001\), so estimated probabilities are truncated to \([0.001,0.999]\). No additional truncation is applied to the clever covariates. The direct Riesz-regression benchmark uses ridge penalty \(\Lambda_{\rm ridge}=1.0\).

\paragraph{Data-generating process.}
For each repetition, generate i.i.d. observations
\[
O=(W_1,W_2,B_1,\ldots,B_8,C_1,\ldots,C_8,A,Z_1,Z_2,R,Y^\dagger),
\]
with latent full outcome \(Y\). The baseline covariates are generated as
\[
W_j\sim N(0,0.5),\qquad j\in\{1,2\},
\]
where the second argument is the variance. The treatment and selection instruments are generated independently as
\[
B_j\sim N(0,0.25),\qquad C_j\sim N(0,0.25),\qquad j\in\{1,\ldots,8\}.
\]
Let \(U_A\sim N(0,2.6/9)\). Treatment is assigned by
\[
A=\mathbb I\{g_A(W,B)+U_A>0\},
\]
where $g_A(W,B)$ is given by
\[
1.15\left\{
0.90W_1-0.10\operatorname{sgn}(W_1)W_1^2
+0.90W_2-0.10\operatorname{sgn}(W_2)W_2^2
+\sum_{j=1}^8 B_j
\right\}.
\]
The post-treatment covariates are
\[
Z_1=-0.50+A+U_{Z_1},
\qquad U_{Z_1}\sim N(0,1),
\]
and, writing \(s_A=2A-1\),
\[
Z_2=0.20\{4.0+0.05s_A+0.50Z_1+0.05s_AZ_1+U_{Z_2}\}^2,
\qquad U_{Z_2}\sim N(0,1).
\]
Let \(U_R\sim N(0,2.6/9)\). Selection is generated by
\[
R=\mathbb I\{g_R(W,A,Z,C)+U_R>0\},
\]
where
\begin{align*}
g_R(W,A,Z,C) &=
-0.90+0.29Z_1+0.54Z_2+\sum_{j=1}^8 C_j
+0.60\sin(W_1+Z_1)\\
&\quad +0.50\tanh\!\left\{0.15\,|W_2|^{Z_2}\operatorname{sgn}(Z_1-W_1)\right\}-1.50\, \mathbb{I}(W_2 < 0).
\end{align*}
The full outcome is
\begin{align*}
Y={}&3.00(W_1+W_2)-0.25s_A+0.50s_A(W_1+W_2)
+1.25Z_1\\
&+0.25s_AZ_1+Z_2+0.50s_AZ_2+U_Y,
\qquad U_Y\sim N(0,7).
\end{align*}
The observed outcome is \(Y^\dagger=Y\) if \(R=1\) and is missing otherwise. The variables \((W_1,W_2)\) are baseline confounders. The variables \((Z_1,Z_2)\) are post-treatment predictors of selection and outcome and are part of the additional adjustment set for the recovered estimand. The variables \((B_1,\ldots,B_8)\) act as treatment instruments, while \((C_1,\ldots,C_8)\) act as selection instruments.

\paragraph{Target parameter and oracle truth.}
The target parameter is the recovered average treatment effect
\[
\psi^\circ=\E^\circ [Q_2^\circ(W,1)-Q_2^\circ(W,0)],
\]
where
\begin{align*}
Q_1^\circ(w,a,z) &=\E^\circ[Y\mid W=w,A=a,Z=z,R=1],\\
Q_2^\circ(w,a) &=\E^\circ[Q_1^\circ(W,A,Z)\mid W=w,A=a].
\end{align*}
The script computes the oracle target analytically from the conditional means of \(Z_1\) and \(Z_2\) under \(A=a\). Specifically,
\[
\mu_{1,a}:=\E^\circ[Z_1\mid A=a]=-0.50+a,
\]
and, with \(s_a=2a-1\) and $\mu_{2,a}:=\E^\circ[Z_2\mid A=a]$,
\[
\mu_{2,a}
=
0.20\left[
(0.50+0.05s_a)^2+1+
\{4.0+0.05s_a+(0.50+0.05s_a)\mu_{1,a}\}^2
\right].
\]
Since \(\E^\circ[W_1+W_2]=0\), the arm-specific mean is
\[
\mu_a
=
-0.25s_a
+1.25\mu_{1,a}
+0.25s_a\mu_{1,a}
+\mu_{2,a}
+0.50s_a\mu_{2,a},
\]
and the oracle value used in all Monte Carlo summaries is
\[
\psi^\circ=\mu_1-\mu_0=5.244625.
\]

\paragraph{Common cross-fitting and engineered features.}
All estimators use the same balanced outer validation-fold indexes. In each repetition, the data are partitioned into \(K=4\) outer folds. For each outer fold, nuisance functions are fit on the corresponding training sample and the fold-specific contribution to the point estimate is computed only on the held-out validation fold. The final estimate is the average over the pooled validation-fold contributions.

Before fitting the nuisance regressions, the script adds deterministic interaction columns \(A_c=Ac\) for
\[
c\in\{W_1,W_2,Z_1,Z_2,B_1,\ldots,B_8,C_1,\ldots,C_8\}.
\]
These engineered features are then used selectively by the deliberately misspecified linear outcome regressions and by the direct Riesz benchmark.

\paragraph{Shared misspecified sequential outcome regressions.}
The plug-in, one-step, TMLE, ridge Riesz, and collaborative estimators all start from the same cross-fitted sequential outcome regressions within each outer training fold. The first-stage regression \(Q_1\) is fit by ordinary least squares among complete cases \(R=1\), using \(Y^\dagger\) as outcome and the covariates
\[
A,W_1,W_2,Z_1,Z_2,B_1,\ldots,B_8,C_1,\ldots,C_8,
AB_1,\ldots,AB_8,AC_1,\ldots,AC_8.
\]
The \(A\)-by-\(W\) and \(A\)-by-\(Z\) interactions are intentionally excluded from this \(Q_1\) regression. The second-stage regression \(Q_2\) is fit by ordinary least squares on the full outer training fold, using the fitted \(\widehat Q_1\) values as pseudo-outcomes and the covariates
\[
A,W_1,W_2,B_1,\ldots,B_8,AB_1,\ldots,AB_8.
\]
The \(A\)-by-\(W\) interactions are intentionally excluded from this \(Q_2\) regression. Thus all non-oracle estimators share the same misspecified linear \(Q_1\) and \(Q_2\) predictions; the estimators differ only in how they use propensity scores, targeting, Riesz corrections, and collaborative representation learning.

\paragraph{Clever covariates and validation-fold targeting.}
For estimated treatment and selection probabilities \(\widehat\pi_A\) and \(\widehat\pi_R\), define
\[
\widehat \alpha_2(O)=
\frac{A}{\widehat\pi_A(W)}-\frac{1-A}{1-\widehat\pi_A(W)},
\qquad
\widehat \alpha_1(O)=
\frac{R\widehat\alpha_2(O)}{\widehat\pi_R(X)},
\qquad
\widehat H_1(X,A)=
\frac{\widehat\alpha_2(O)}{\widehat\pi_R(X)},
\]
where \(X=(W,A,Z)\), so \(\widehat\alpha_1=R\widehat H_1\). 

The two-stage additive targeting routine first estimates \(\epsilon_1\) by least squares on the complete cases in the relevant fold,
\[
\epsilon_1
=
\frac{\sum_{i:R_i=1} \widehat H_{1,i}\{Y_i^\dagger-\widehat Q_1(X_i)\}}
{\sum_{i:R_i=1} \widehat H_{1,i}^2}.
\]
It then updates
\[
\widehat Q_1^*(X_i)=\widehat Q_1(X_i)+\epsilon_1\widehat H_{1,i}.
\]
Second, it estimates
\[
\epsilon_2
=
\frac{\sum_i \widehat \alpha_{2,i}\{\widehat Q_1^*(X_i)-\widehat Q_2(W_i,A_i)\}}
{\sum_i \widehat \alpha_{2,i}^2},
\]
and updates
\[
\widehat Q_2^*(W,1)=\widehat Q_2(W,1)+\frac{\epsilon_2}{\widehat\pi_A(W)},
\qquad
\widehat Q_2^*(W,0)=\widehat Q_2(W,0)-\frac{\epsilon_2}{1-\widehat\pi_A(W)}.
\]
The validation-fold TMLE contribution is \(\widehat Q_2^*(W,1)-\widehat Q_2^*(W,0)\). This same final validation-fold targeting routine is used for the non-collaborative TMLE benchmark and for the selected collaborative candidate.

\paragraph{Non-collaborative benchmark estimators.}
The simulation reports four non-collaborative benchmark estimators and one oracle-style diagnostic:
\begin{itemize}
\item \texttt{linearQ\_plugin}, the cross-fitted plug-in estimator \(\E_n[\widehat Q_2(W,1)-\widehat Q_2(W,0)]\);
\item \texttt{linearQ\_neuralG\_one\_step}, the cross-fitted one-step estimator using the shared linear \(Q_1,Q_2\) predictions and non-collaborative neural treatment and selection propensities;
\item \texttt{linearQ\_neuralG\_tmle}, the cross-fitted TMLE benchmark using the same shared linear \(Q_1,Q_2\) predictions and the same non-collaborative neural propensity architecture;
\item \texttt{linearQ\_ridge\_riesz}, a direct ridge Riesz-regression benchmark using the shared linear \(Q_1,Q_2\) predictions;
\item \texttt{oracle\_nuisance\_one\_step}, a diagnostic one-step estimator evaluated on the same outer validation folds using analytic nuisance-function helpers from the simulation script.
\end{itemize}

The non-collaborative treatment propensity model uses inputs
\[
W_1,W_2,B_1,\ldots,B_8,
\]
and the non-collaborative selection propensity model uses inputs
\[
A,W_1,W_2,Z_1,Z_2,C_1,\ldots,C_8.
\]
These propensity networks have the same encoder and head architecture as the collaborative propensity networks, with bottleneck dimension \(d=6\) for the non-collaborative benchmarks, but their auxiliary drift losses and bottleneck penalties are set to zero.

The one-step benchmark evaluates, on each outer validation fold,
\[
\widehat Q_2(W,1)-\widehat Q_2(W,0)
+\widehat \alpha_1(O)\{Y^\dagger-\widehat Q_1(X)\}
+\widehat \alpha_2(O)\{\widehat Q_1(X)-\widehat Q_2(W,A)\}.
\]
The TMLE benchmark applies the validation-fold targeting step described above and averages the resulting updated contrast over the pooled validation folds.

\paragraph{Direct ridge Riesz-regression benchmark.}
The direct Riesz benchmark approximates the two residual weights by ridge regressions on hand-crafted bases. For the term multiplying \(R(Y^\dagger-\widehat Q_1)\), the ridge regression is trained among complete cases to approximate \(\widehat H_1\) using the main effects
\[
A,W_1,W_2,Z_1,Z_2,B_1,\ldots,B_8,C_1,\ldots,C_8
\]
and all interactions between \(A\) and each non-\(A\) covariate in this list. For the term multiplying \(\widehat Q_1-\widehat Q_2\), the ridge regression is trained on the full outer training fold to approximate \(\widehat \alpha_2\) using the main effects
\[
A,W_1,W_2,B_1,\ldots,B_8
\]
and all interactions between \(A\) and each non-\(A\) covariate in this list. Each ridge regression is implemented with feature standardization followed by ridge regression with penalty \(\Lambda_{\rm ridge}=1.0\). On a validation fold, the Riesz contribution is
\[
\widehat Q_2(W,1)-\widehat Q_2(W,0)
+R\widehat H_1^\star(X,A)\{Y^\dagger-\widehat Q_1(X)\}
+\widehat\alpha_2^\star(O)\{\widehat Q_1(X)-\widehat Q_2(W,A)\},
\]
where \(\widehat H_1^\star\) and \(\widehat\alpha_2^\star\) are the fitted ridge representer approximations.

\paragraph{Collaborative representation CTMLE.}
The collaborative estimators use the same shared linear \(Q_1,Q_2\) predictions as the benchmark estimators, but learn treatment and selection propensities through collaborative representation networks along a budgeted path. For treatment, the network takes inputs
\[
W_1,W_2,B_1,\ldots,B_8
\]
and maps them through a two-hidden-layer ReLU encoder \(h_A\) into a \(d\)-dimensional code. A propensity head maps this code to \(\widehat\pi_A(W)\), and a two-output auxiliary head predicts the factual-arm treatment drift proxy. For selection, the network takes inputs
\[
W_1,W_2,A,Z_1,Z_2,C_1,\ldots,C_8,
\]
maps them through a two-hidden-layer ReLU encoder \(h_R\), and uses a propensity head with a direct skip connection for \((W_1,W_2,A)\). The selection auxiliary head predicts the complete-case selection drift proxy. All neural propensity components use hidden width \(64\).

For a fixed library element, the budget path is indexed by \(b\in\{0,\ldots,B\}\), with \(B=3\). Candidate \(b=0\) uses the shared linear sequential outcome regressions together with constant treatment and selection probabilities equal to the clipped training-fold marginal probabilities \(\E_n[A]\) and \(\E_n[R]\). This endpoint is retained as the initialization of the path, but is not a selectable final collaborative estimator. For \(1\leq b<B\), the current candidate defines
\[
e_R^b=\widehat Q_1^b(X)-Y^\dagger\qquad (R=1),
\]
\[
\xi^b=
\widehat Q_1^b(X)+
\frac{R}{\widehat\pi_R^{b-1}(X)}\{Y^\dagger-\widehat Q_1^b(X)\},
\]
and
\[
e_A^b=\widehat Q_2^b(W,A)-\xi^b.
\]
The selection network is trained with binary cross-entropy for \(R\), the complete-case auxiliary squared loss for \(e_R^b\), and a bottleneck penalty on \(h_R\). The treatment network is trained with binary cross-entropy for \(A\), a weighted factual-arm squared loss for \(e_A^b\), and a bottleneck penalty on \(h_A\). The treatment auxiliary weights are
\[
\frac{A}{\widehat\pi_A^{b-1}(W)}
+
\frac{1-A}{1-\widehat\pi_A^{b-1}(W)},
\]
with probabilities clipped at \(\epsilon_{\rm clip}=0.001\). After the collaborative propensities are fit at an intermediate budget level, the script applies the two-stage TMLE fluctuation on the training fold and carries the resulting updates to the corresponding validation fold.

At the terminal budget \(b=B\), the collaborative auxiliary losses and bottleneck penalties are zero. In the implementation, this endpoint is made directly comparable to the non-collaborative neural-propensity TMLE benchmark by fitting ordinary propensity networks with the same architecture and no collaborative penalties. Thus \(b=B\) is a neural-propensity TMLE-compatible fallback, while \(b=0\) is a constant-propensity initialization.

The budget-dependent collaborative penalties are
\[
\lambda_A^b=\lambda_A(1-b/B)^\rho,
\quad
\lambda_R^b=\lambda_R(1-b/B)^\rho,
\quad
\tau_A^b=\tau_A(1-b/B)^\rho,
\quad
\tau_R^b=\tau_R(1-b/B)^\rho.
\]
The reported collaborative implementations are
\begin{itemize}
\item the \(\theta_1\)-only singleton library with
\[
(d_A,d_R,\omega,\lambda_A,\lambda_R,\tau_A,\tau_R,\rho)
=(3,3,0.6,2.00,5.00,0.015,0.005,1.0);
\]
\item the \(\theta_2\)-only singleton library with
\[
(d_A,d_R,\omega,\lambda_A,\lambda_R,\tau_A,\tau_R,\rho)
=(4,4,0.8,3.00,3.00,0.010,0.010,1.0);
\]
\item the two-element library \(\Theta=\{\theta_1,\theta_2\}\).
\end{itemize}

\paragraph{Inner targeted selection and final validation-fold evaluation.}
For each outer training fold, the collaborative estimator performs two-fold inner cross-validation. For every library element and every budget level \(b\in\{0,\ldots,B\}\), the script fits the candidate path on the inner training portion and scores each candidate on the corresponding inner validation portion before any additional validation-fold targeting. This pre-targeting score avoids making the empirical influence-function mean nearly zero for every candidate by construction. Candidate \(b=0\) is included as a path initialization but is excluded from the final inner-selection search, so only \(b\in\{1,\ldots,B\}\) can be selected.

For a candidate \(s\), the inner targeted risk is
\[
\widehat r(s)
=
\widehat b(s)^2+\frac{1}{n_{\mathrm{train}}}\,\widehat v(s)
+\zeta\widehat\ell_{\omega}(s),
\]
where \(\widehat b(s)\) is the inner-validation mean of the estimated influence-function value, \(\widehat v(s)\) is the corresponding variance estimate, and
\[
\widehat \ell_{\omega}(s)
=
\E_n\{R(Y^\dagger-\widehat Q_1)^2\}
+
\omega\,\E_n\{(\widehat Q_1-\widehat Q_2(W,A))^2\}
\]
is an optional sequential regression-loss stabilizer. In the reported run, \(\zeta=0\), so the sequential-loss term is not used in the selector. The selected library element and budget are refit on the full outer training fold, the resulting selected candidate is targeted on the outer validation fold, and the updated validation-fold contrast is pooled over outer folds.


%
%

\end{document}